\documentclass{iopjournal-arxiv}

\usepackage{cite}
\usepackage{fancyhdr}
\usepackage{hyperref}
\usepackage{amsmath,amssymb,amsfonts}
\usepackage{algorithmic}
\usepackage{algorithm}
\usepackage{graphicx}
\usepackage{textcomp}
\usepackage{xcolor}
\usepackage{subcaption}
\usepackage{orcidlink}

\begin{document}

\title{VQA for Dynamic Portfolio Optimization: Sampling Strategies, Optimizer Scheduling, and Hardware-Aware Ansatz Design}

\author{Mohammad Kashfi Haghighi\orcid{0009-0008-6263-7900}}
\affil{PINQ\textsuperscript{2}, Montreal, Canada}


\begin{abstract}
Variational quantum algorithms are being increasingly explored for optimization problems at scales relevant to near-term quantum devices. Their practical performance, however, depends strongly on design choices such as the sampling objective, classical optimizer, and layout of the parameterized ansatz before and after hardware transpilation. In this work, we study these factors in the context of dynamic portfolio optimization, a multi-period financial optimization problem that balances return, risk, transaction costs, cash-interest effects, and portfolio constraints. We use a sampling-based VQA framework on a 150-qubit dynamic portfolio instance and evaluate several components of the optimization workflow. We propose a specific adaptive CVaR schedule that gradually tightens the sampled tail used for optimization, together with a two-stage classical optimizer combining global exploration with Particle Swarm Optimization and local refinement with the Nakanishi-Fujii-Todo optimizer. We also study ansatz depth and sequential growth strategies. Finally, we introduce two hardware-aware ansatz-layout modifications: a data-guided colored layout that assigns correlated variables to qubits connected by entangling gates, and a heavy-hex-native deep-chain layout designed to increase native two-qubit interaction depth without additional routing overhead after transpilation. Simulator studies are used to select CVaR, optimizer, and depth configurations, while the final ansatz comparison is performed on the \texttt{ibm\_quebec} QPU. The results show that sampling strategy, optimizer scheduling, and hardware-aware layout design materially affect performance. In the reported QPU layout comparison, the proposed heavy-hex-native deep-chain layout achieves the best final objective value and CVaR-tail performance among the tested layouts. Although we do not observe quantum advantage over a state-of-the-art exact classical solver, our results provide practical guidance for improving VQA performance on near-term hardware.
\end{abstract}

\section{Introduction}
Recent progress in quantum hardware, compilation, and algorithms has enabled experiments beyond small toy instances and toward larger benchmark instances motivated by industrial optimization problems \cite{kim2023evidence, van2023probabilistic, angara2026experimental}. In particular, noisy intermediate-scale quantum (NISQ) devices \cite{Preskill2018NISQ} now support circuits involving more than one hundred qubits, which has motivated renewed interest in using current quantum processors for optimization problems arising in finance and chemistry \cite{Agliardi2025SamplingVQA, Pelofske2024ScalingWholeChipQAOA, Abbas2024QuantumOptimization}. However, realizing consistent algorithmic gains from these devices remains challenging. Noise, finite sampling budgets, and transpilation overhead all affect the practical performance of quantum algorithms. Therefore, algorithmic design, ansatz construction, optimizer selection, and hardware-aware compilation remain critical for making effective use of current quantum processors \cite{leone2024practical, larocca2025barren, BonetMonroig2023PerformanceComparison, yuan2025full, minor-embedding}.

Among the most widely studied approaches for NISQ optimization are Variational Quantum Algorithms (VQAs) \cite{Cerezo2021VQA}, including the Quantum Approximate Optimization Algorithm (QAOA) \cite{farhi2014quantum, blekos2024review} and the Variational Quantum Eigensolver (VQE) \cite{Peruzzo2014VQE, tilly2022variational}. In contrast to fault-tolerant algorithms such as Shor's \cite{shor1994algorithms} or Grover's \cite{Grover1996FastQuantumSearch} algorithm, which require deep circuits to achieve their asymptotic advantages, VQAs are designed around shallow parameterized circuits.
A quantum processor is used to prepare and sample a variational circuit, while a classical optimizer updates the circuit parameters using objective values computed from the sampled bitstrings. This hybrid feedback loop is well suited to NISQ devices because a hardware-feasible ansatz can be chosen in advance, while the classical optimizer tunes its parameters, making VQAs a natural candidate for large-scale heuristic optimization.

Portfolio optimization is a prominent application area for quantum optimization. In its basic form, the goal is to select or weight assets so as to balance expected return and risk. Because realistic formulations often involve discrete allocation decisions, cardinality limits, transaction costs, and budget constraints, the resulting optimization problem can become difficult at large scale \cite{koch2025quantum}. 
Cardinality constraints are especially important in realistic portfolio construction because they model limited active positions, but they also introduce combinatorial structure that makes the resulting mean--variance portfolio selection problem NP-hard in general \cite{GaoLi2013CardinalityPortfolio}.

Dynamic portfolio optimization is an even richer formulation: instead of choosing a single static portfolio, the investor chooses a sequence of
portfolio allocations over multiple rebalancing periods. This introduces time-coupled terms, such as transaction costs and liquidation costs, and makes the optimization problem more challenging than a single-period allocation problem \cite{skaf2009multi}. These characteristics make dynamic portfolio optimization a suitable testbed for studying the performance of VQA-based optimization methods.

Several recent works have investigated VQA for portfolio optimization \cite{buonaiuto2023best, Agliardi2025SamplingVQA, Nodar2025ScalingVQE}.
However, there is still substantial room for improvement in the design of the objective aggregation strategy, the classical optimizer, and the ansatz layout. One important direction is the use of Conditional Value-at-Risk (CVaR) aggregation, where the optimizer focuses on the low-cost tail of the sampled distribution rather than
on the average over all samples \cite{Barkoutsos2020CVaR}. CVaR-based objectives can be especially useful in combinatorial optimization, where the practical goal is often to increase the probability of sampling high-quality solutions. Another important direction is optimizer selection \cite{mcclean2018barren, larocca2025barren, BonetMonroig2023PerformanceComparison}. Variational landscapes are typically noisy, non-convex, and high-dimensional, and the performance of a VQA run can depend strongly on whether the optimizer is better suited for global exploration or local refinement.

Ansatz depth creates a tradeoff between expressivity, trainability, and hardware cost \cite{sim2019expressibility, Holmes2022ExpressibilityTrainability, leone2024practical}. This tradeoff is further complicated on noisy hardware, where increasing depth can also worsen trainability through noise-induced barren plateaus \cite{Wang2021NoiseInducedBarrenPlateaus}.
Increasing the number of entangling operations can improve the circuit's ability to represent correlated bitstring distributions \cite{sim2019expressibility}, but the way depth is added matters. 
Repeating an ansatz block typically increases the number of variational parameters, which can make optimization harder under a fixed evaluation budget. Alternatively, adding more two-qubit interactions within a layer can increase two-qubit interaction depth without necessarily increasing the parameter count, but it may also introduce additional transpilation overhead on the target hardware.

This hardware constraint is important on current quantum processors including IBM superconducting processors, such as Heron-generation devices, which use sparse heavy-hex-style connectivity rather than all-to-all connectivity \cite{chamberland2020topological}. If an ansatz requests two-qubit interactions that cannot be embedded directly on adjacent physical qubits, routing passes such as SABRE may insert SWAP operations to satisfy the coupling map, increasing both the two-qubit gate count and the transpiled depth \cite{Li2019SABRE}. Thus, a useful ansatz for current hardware should not only be expressive at the logical level, but should also be designed so that its two-qubit interactions can be embedded on the hardware graph with limited routing overhead.

Inspired by the framework in \cite{Agliardi2025SamplingVQA}, we study sampling-based VQA for a 150-qubit dynamic portfolio optimization instance and propose improvements to three parts of the VQA workflow: CVaR scheduling, classical optimization, and ansatz layout design. First, we introduce an adaptive CVaR schedule. Instead of fixing the CVaR level
throughout training, the schedule starts from a broader low-cost tail and gradually tightens the CVaR level as optimization progresses. The motivation is that early iterations should provide enough information to move the circuit toward feasible or near-feasible regions, while later iterations can focus more strongly on the best sampled portfolios. Our experiments show that the adaptive schedule is competitive with fixed CVaR baselines and can improve the final low-tail sample quality while reducing the average shot cost compared with using the fixed CVaR levels throughout training.
Related work has also explored changing CVaR-like objectives during variational optimization \cite{Kolotouros2022EvolvingCVaR}; our schedule differs in moving in the opposite direction: it decreases \(\alpha\) during training as an exploration-to-refinement mechanism for constrained sampling-based DPO.

Second, we propose a two-stage classical optimizer schedule. The optimizer begins with Particle Swarm Optimization (PSO) \cite{Kennedy1995PSO}, which provides global exploration of the parameter space, and then switches to the Nakanishi-Fujii-Todo (NFT) \cite{Nakanishi2020NFT} optimizer for local coordinate-wise refinement. The switch is controlled by a minimum PSO budget, a maximum PSO budget, and a stagnation criterion. This design is intended to combine the exploration capability of PSO with the local refinement behavior of NFT.

Third, we introduce two ansatz-layout modifications. The first is a Data-Guided Colored Layout (DGC), which keeps a hardware-compatible colored entangling schedule but modifies the mapping from portfolio variables to circuit qubits. The mapping is chosen so that variables with stronger empirical dependence, estimated through a mutual-information-based affinity matrix, are more likely to be placed on qubits connected by entangling edges. This adds a problem-dependent inductive bias without increasing the two-qubit depth of the colored layout. The second is a Heavy-Hex Native Deep-Chain (HNDC) layout, which is designed to increase the native two-qubit interaction depth within a single ansatz repetition while using only hardware-compatible couplings. The goal is to enrich the circuit’s native two-qubit connectivity structure without introducing the large routing overhead that would arise from non-native long-range interactions.

We evaluate these components in two stages. First, we perform simulator-based studies to select the CVaR strategy, optimizer schedule, and ansatz repetition strategy under a controlled experimental protocol. Then, using the selected configuration, we compare the ansatz layouts on the \texttt{ibm\_quebec} QPU. The results show that hardware-aware layout design has a substantial effect on performance: the proposed HNDC layout achieves the best performance among the tested layouts in the hardware study, while the DGC layout improves over the standard colored layout with the same two-qubit depth.

The remainder of this paper is organized as follows. Section~\ref{sec:background} introduces the dynamic portfolio optimization formulation and the sampling-based VQA framework used in this work. Section~\ref{sec:methodology} describes the proposed methods, including adaptive CVaR scheduling, the two-stage optimizer schedule, sequential depth growth, and ansatz layout design. Section~\ref{sec:experiments} presents the simulator and QPU experiments. Section~\ref{sec:discussion} discusses the main findings and limitations. Finally, Section~\ref{sec:conclusion} concludes the paper and outlines directions for future work.

\section{Background}
\label{sec:background}

\subsection{Dynamic Portfolio Optimization}
\label{sec:dpo}

We adopt a multi-period (dynamic) portfolio model inspired by the portfolio benchmarking formulation in the Quantum Optimization Benchmark Library \cite{koch2025quantum}, while keeping the instance size compatible with near-term quantum optimization experiments. The model considers \(n\) assets traded over \(m\) discrete rebalancing periods \(t=0,\dots,m-1\). Short selling is not allowed.

\paragraph{Data and decision variables}
Let \(p_{i,t}\) be the (anchor) price of asset \(i\in\{1,\dots,n\}\) at the beginning of period \(t\), with \(p_{i,t+1}\) the corresponding price at the next anchor. Let \(\Sigma_t\in\mathbb{R}^{n\times n}\) denote the covariance matrix of \emph{percentage returns} at period \(t\) (used only within-period, with no cross-time covariance terms). The decision variable is the integer number of shares held:
\[
h_{i,t}\in\mathbb{Z}_{\ge 0},\qquad i=1,\dots,n,\;\; t=0,\dots,m-1.
\]
To obtain a binary optimization problem, we encode each integer holding using \(K\) binary bits:
\[
x_{i,t,k}\in\{0,1\},\quad
h_{i,t}=\sum_{k=0}^{K-1}2^k\,x_{i,t,k}.
\]
In the implementation we allow \(h_{i,t}=0\) (asset not held), and restrict non-zero holdings to a maximum of \(h_{\max}\) via a soft penalty (see below).

\paragraph{Objective}
We minimize a sum of per-period costs capturing risk, realized price change, transaction costs, and a cash interest term.
Let \(h_t\in\mathbb{R}^n\) be the holdings vector at time \(t\), and define the dollar exposure \(v_t = p_t \odot h_t\), where \(\odot\) denotes elementwise multiplication and \(p_t=(p_{1,t},\dots,p_{n,t})\).
The objective used in our experiments is:
\begin{align}
\min_{x}\;\; &\sum_{t=0}^{m-1} \Big[
\underbrace{\lambda \, v_t^\top \Sigma_t v_t}_{\text{risk (dollar-weighted variance)}}
\;-\;\underbrace{\sum_{i=1}^n (p_{i,t+1}-p_{i,t})\,h_{i,t}}_{\text{return (price change)}} \nonumber\\
&\hspace{.5cm}
+\;\underbrace{\delta \sum_{i=1}^n p_{i,t}\,\lvert h_{i,t}-h_{i,t-1}\rvert}_{\text{transaction cost}}
\;-\;\underbrace{\nu (W-\sum_{i=1}^n p_{i,t}h_{i,t})}_{\text{cash interest term}}
\Big]  \label{eq:dpo_obj}\\
&\quad+\;\underbrace{\delta\sum_{i=1}^n p_{i,m}\,h_{i,m-1}}_{\text{liquidation cost at terminal anchor}},
\nonumber
\end{align}
with the convention \(h_{i,-1}=0\) (starting from an all-cash portfolio). Here \(\lambda\ge 0\) is the risk-aversion coefficient, \(\delta\ge 0\) scales transaction and liquidation costs, \(\nu\ge 0\) is the cash interest rate, and \(W\) is the initial cash budget. Let \(V_t=\sum_{i=1}^n p_{i,t}h_{i,t}\) denote the total invested value at time \(t\). The term \(-\nu(W-V_t)\) models the interest benefit of uninvested cash \(W-V_t\). Since the constant \(-\nu W\) does not affect optimization, the term is equivalent to \(+\nu V_t\) up to an additive constant; it therefore represents the opportunity cost of deploying capital, and can be interpreted as a borrowing cost if \(V_t>W\).

\paragraph{Cardinality (active-asset) control.}
To reflect practical portfolio-construction settings with a limited target number of active positions, we control the number of \emph{active} assets per period using a soft penalty.
Let
\[
a_t = \sum_{i=1}^n \mathbb{I}[h_{i,t}>0]
\]
be the number of assets with non-zero holdings at time \(t\). In the implementation, we penalize deviations from a target cardinality \(B\) using
\[
P_{\mathrm{card}} \sum_{t=0}^{m-1} \lvert a_t - B\rvert,
\]
which encourages \(B\) active assets each period.

\paragraph{Holding-validity penalty}
To enforce an upper bound on position size, we penalize active-asset holdings that exceed \(h_{\max}\).
\[
P_{\mathrm{hold}}\sum_{t=0}^{m-1}\sum_{i=1}^n 
\begin{cases}
0,           & h_{i,t}\le h_{\max}, \\
h_{i,t}-h_{\max}, & h_{i,t}>h_{\max}.\\
\end{cases}
\]

\subsection{(Sampling-Based) VQA}
\label{sec:vqe_background}

A key practical limitation of many gate-based quantum algorithms on current NISQ devices is the required depth in two-qubit operations, since deeper circuits generally require longer coherent evolution and are therefore more susceptible to decoherence and accumulated gate noise. As a result, hybrid variational approaches have become a common route for leveraging current quantum hardware in optimization tasks. In VQAs, a parameterized circuit \(U(\boldsymbol{\theta})\) prepares a trial state \(|\psi(\boldsymbol{\theta})\rangle = U(\boldsymbol{\theta})|0\rangle\). Measurements in the computational basis produce bitstrings \(x\), whose objective values are evaluated classically. A classical optimizer then updates \(\boldsymbol{\theta}\), and the process is repeated until a stopping criterion is met.

For combinatorial optimization problems, the objective is diagonal in the computational basis, so each measured bitstring corresponds to a classical cost value \(f(x)\). The standard VQA objective uses the sample mean as an estimator of the expected energy. However, for classical optimization, the practical goal is often to increase the probability of sampling high-quality solutions rather than to optimize the average value of all observed samples. In \cite{Barkoutsos2020CVaR}, CVaR is proposed as an alternative aggregation rule for variational quantum optimization. Given \(N\) samples \(\{x^{(s)}\}_{s=1}^N\), sorted by increasing objective value, the empirical CVaR objective at level \(\alpha\in(0,1]\) is
\begin{equation}
\label{eq:cvar_estimator}
\mathrm{CVaR}_{\alpha}
=
\frac{1}{\lceil \alpha N\rceil}
\sum_{s=1}^{\lceil \alpha N\rceil}
f\!\left(x^{(s)}\right).
\end{equation}
Thus, \(\alpha=1\) recovers the usual sample mean, while smaller values of \(\alpha\) focus the optimizer on the lower-cost tail of the sampled distribution. This makes CVaR a useful objective for variational optimization when the goal is to bias the circuit toward better candidate solutions.

VQA has already been studied for portfolio optimization \cite{Nodar2025ScalingVQE, Agliardi2025SamplingVQA, buonaiuto2023best, wang2025variational}. In \cite{Nodar2025ScalingVQE}, the DPO objective function is formulated based on Sharpe ratio and transformed into a QUBO and an Ising Hamiltonian, which is optimized using a standard VQA workflow. That work systematically studies how VQA scales from small instances to problems above 100 qubits, and shows that the combination of ansatz and classical optimizer has a strong effect on performance. In particular, the best results are obtained using Differential Evolution together with an ansatz tailored to both the problem structure and the QPU connectivity.

A different approach is the sampling-based CVaR-VQA workflow used for bond portfolio construction in \cite{Agliardi2025SamplingVQA}. In this setting, the circuit acts as a trainable sampler: measured bitstrings are evaluated directly by a classical objective function, rather than through an explicit QUBO-to-Ising Hamiltonian construction. This is useful when the objective includes terms that are inconvenient to represent quadratically, such as absolute-value transaction costs, nonlinear penalties, or inequality constraints that would otherwise require slack variables.

Constraint-aware ansatz constructions have also been studied in variational quantum optimization, particularly through QAOA and the quantum alternating-operator ansatz \cite{Hadfield2019QAOAOperatorAnsatz}. However, enforcing feasibility directly at the circuit level often requires problem-specific constraint-preserving mixers or additional operations, which can increase circuit depth and reduce portability across problem formulations. In this work, we take a different route and follow the sampling-based setting with soft penalties: at each iteration, bitstrings sampled from \(U(\boldsymbol{\theta})\) are decoded into integer holdings, and the dynamic portfolio objective is evaluated directly on those holdings. This retains the flexibility of direct objective evaluation while still allowing CVaR-based optimization of the sampled objective values.

In \cite{Nodar2025ScalingVQE}, two hardware-efficient ansatz layouts are considered for dynamic portfolio optimization: a bilinear layout and a colored layout. The bilinear layout follows a regular nearest-neighbor entanglement pattern, while the colored layout is designed to better exploit the connectivity of the target quantum hardware. We use these layouts as baselines and describe them in more detail in Section~\ref{sec:ansatz_layouts}. These prior approaches motivate the layout designs and broader VQA workflow choices described in Section~\ref{sec:methodology}.

\subsection{Correlation- and Entanglement-Informed Ansatz Design}
\label{sec:entangled_ansatz}
Our data-guided colored layout (DGC, introduced in Section~\ref{sec:dgc_layout}) is related in spirit to correlation- and entanglement-informed ansatz design, where qubit placement or circuit connectivity is adapted to structural information in the target problem. PermVQE~\cite{tkachenko2021correlation} uses mutual-information-based qubit permutations to place strongly correlated spin-orbitals near one another, reducing the circuit depth required in molecular VQE. Joch et al.~\cite{joch2025entanglement} show that tailoring the entangling-gate structure of VQE ansätze to the expected entanglement pattern of quasi-1D ground states can reduce the resources needed to reach a target accuracy. A portfolio-specific approach is EAQGA~\cite{haghighi2025eaqga}, which combines portfolio covariance information with a genetic-algorithm search to identify highly correlated variables and preferentially entangle them during circuit construction.
DGC addresses a different design question. Rather than permuting qubits to reduce depth or searching over entanglement patterns, it keeps a fixed hardware-native colored entangling schedule and uses empirical asset correlations only to assign portfolio variables to the qubit pairs that schedule already entangles. In this sense it couples portfolio-derived correlation information directly with the device's native connectivity, without any search over entanglement structure and without adding two-qubit depth.

\section{Methodology}\label{sec:methodology}
\subsection{Adaptive CVaR Schedule}
\label{sec:adaptive_cvar}
Schedules that vary the CVaR level during variational optimization have been explored previously. The Ascending-CVaR method of Kolotouros and Wallden~\cite{Kolotouros2022EvolvingCVaR} starts with a small \(\alpha\) and progressively widens it toward the full expectation value (\(\alpha=1\)), with the motivation that the location of local minima changes with \(\alpha\) while the global minimum remains invariant for \(\alpha\) below the maximum reachable overlap, so widening can free the optimizer from sub-optimal basins. Our adaptive schedule moves in the opposite direction: it starts with a wider \(\alpha\) to provide a broad training signal that drives the circuit toward feasible regions, and gradually tightens \(\alpha\) as feasibility improves so that the objective concentrates on high-quality feasible samples. This descending exploration-to-refinement pattern is motivated by the very small feasible fraction of the constrained DPO search space, quantified below.

In our setting, for example, \(n=10\), \(m=5\), \(K=3\), \(B=4\), and \(h_{\max}=5\), so the full binary space contains \(2^{n m K}=2^{150}\) bitstrings. If feasibility is approximated by exactly \(B\) active assets per period and nonzero holdings in \(\{1,\dots,h_{\max}\}\), then the number of feasible configurations is
\[
\left[\binom{10}{4}5^4\right]^5,
\]
which gives a feasible fraction of approximately
\[
\frac{\left[\binom{10}{4}5^4\right]^5}{2^{150}}
\approx 2.73\times 10^{-20}.
\]
Thus, at initialization, a circuit that samples broadly over the search space is unlikely to assign substantial probability mass to feasible high-quality portfolios.

This creates a tradeoff in choosing a fixed CVaR level. If \(\alpha\) is chosen too small from the beginning, the objective may depend on only a few rare low-cost samples. Such samples may be informative, but they can also be accidental and provide an incomplete view of the landscape. In this regime, the optimizer may prematurely chase isolated samples rather than learn parameter directions that systematically increase the probability of feasible portfolios. A larger \(\alpha\) gives a broader training signal and allows the optimizer to account more consistently for the penalty structure of the problem. In this sense, the early phase is mainly an exploration phase whose purpose is to move probability mass toward feasible or near-feasible regions.

As training progresses and the circuit parameters improve, we expect the sampled distribution to contain a higher proportion of feasible or near-feasible portfolios, making a smaller CVaR level more effective for concentrating the optimization on high-quality candidates. Keeping \(\alpha\) too large at this stage can be inefficient: improved feasible samples may be averaged together with many high-penalty infeasible samples, so the aggregate objective may change only weakly even when the best candidates improve. This can lead to an apparent plateau in the optimization. A smaller \(\alpha\) reduces this effect by focusing the objective on the low-cost tail, where feasible and high-quality portfolios are more likely to appear.

To encode this exploration-to-refinement behavior, we use a decreasing CVaR schedule:
\[
\alpha(r)=
\max\left(
\alpha_{\min},
\alpha_{\max}
-
\Delta\alpha
\left\lfloor \frac{r}{L_\alpha}\right\rfloor
\right),
\]
where \(r\) denotes the iteration index. The stepwise form is intentional: changing \(\alpha\) changes the effective objective landscape, so we avoid changing it too frequently inside the inner optimizer dynamics.

In particular, the schedule is updated only at optimizer-block boundaries. For NFT updates, all function evaluations used to update a given parameter are computed with the same CVaR level. This prevents the coordinate-wise update from mixing different objective landscapes while estimating the minimum. The proposed adaptive CVaR schedule is therefore a controlled heuristic: it uses broader tail information early to guide the circuit toward feasibility, and then gradually shifts toward more selective tail-focused optimization to improve the quality of the best feasible samples.

\subsection{Two-Stage Classical Optimizer Schedule (PSO $\rightarrow$ NFT)}
\label{sec:two_stage}

The choice of classical optimizer can strongly affect the performance of variational quantum algorithms, especially when the objective is noisy, non-convex, and high-dimensional. In this work, we compare three standard derivative-free optimizers---COBYLA \cite{Powell1994COBYLA}, Particle Swarm Optimization (PSO) \cite{Kennedy1995PSO}, and the Nakanishi-Fujii-Todo (NFT) \cite{Nakanishi2020NFT} optimizer---against a proposed two-stage classical optimizer schedule.

We use COBYLA as a local derivative-free baseline. COBYLA constructs linear approximations of the objective and constraints inside a trust-region framework, and is commonly used when gradient information is unavailable.
PSO is used as a population-based global-search baseline: a swarm of particles explores the parameter space by combining each particle's best-known position with the best-known position found by the swarm. NFT, in contrast, is a coordinate-wise derivative-free optimizer originally designed for parameterized quantum circuits. It updates one parameter at a time: for a chosen parameter \(\theta_j\), all other parameters are fixed and the objective is evaluated at a small number of shifted points, typically including \(\theta_j\) and \(\theta_j \pm \pi/2\). For expectation-value objectives generated by suitable single-parameter rotation gates, the dependence on \(\theta_j\) has a simple sinusoidal form, allowing NFT to estimate the one-dimensional profile analytically and move \(\theta_j\) to the corresponding minimizer before proceeding to the next parameter. This makes NFT most appropriate as a local refinement method once the parameters are already in a promising region. In our sampling-based setting, this sinusoidal model should be understood as an optimization heuristic rather than an exact property of the empirical objective, because the CVaR aggregation involves finite-shot sampling, sorting, and tail selection.

Motivated by the complementary behavior of PSO and NFT, and following the broader exploration-then-refinement pattern of memetic methods in classical optimization~\cite{neri2012memetic}, we introduce a two-stage schedule. The first stage uses PSO to explore the parameter space and identify a good candidate parameter vector. The second stage initializes NFT from the best parameter vector found by PSO and performs local coordinate-wise refinement. All circuit parameters are kept within the periodic interval \([0,2\pi]\).

The transition from PSO to NFT is adaptive. PSO is assigned a minimum and maximum evaluation budget. After the minimum budget is reached, we monitor the best objective value over a sliding window of evaluations. If the improvement over the window is below a prescribed threshold, the PSO phase is considered stagnant and is stopped
early. The remaining evaluation budget is then transferred to NFT. This prevents excessive evaluations from being spent on an unproductive global search, while still allowing PSO enough time to locate a promising basin before local refinement begins.

This optimizer is used as a practical exploration--refinement strategy for the VQA training loop: PSO provides robustness to initialization, while NFT provides efficient local improvement once a good region has been found.

\subsection{Ansatz Depth Strategy and Sequential Growth}
\label{sec:depth_strategy}

Increasing the number of ansatz repetitions generally increases circuit expressivity \cite{sim2019expressibility}, but it also increases the number of variational parameters and can make the optimization landscape more difficult.

Instead of optimizing a deeper circuit from random initialization, layerwise training strategies use parameters learned at smaller depth to initialize deeper circuits. A related parameter-fixing strategy has been proposed for QAOA, where optimal parameters obtained at lower depth are used when constructing higher-depth circuits \cite{Lee2021ParameterFixing}.

We adapt this idea to the VQA setting and evaluate a sequential depth-growth procedure. The procedure starts with a one-layer ansatz and optimizes its parameters under the chosen VQA configuration. A second layer is then appended, while the parameters of the first layer are kept fixed, and only the newly added layer is optimized. This process can be repeated to construct deeper ansatzes. Since fixing earlier layers may prevent the full circuit from adapting after new layers are added, we include a final joint refinement stage in which all parameters are released and optimized together, initialized from the parameters found during the sequential stages.

We compare this sequential-growth strategy with fixed-depth ansatzes using one, two, and three repetitions.
To ensure a fair comparison, all repetitions configurations are evaluated under the same total objective-evaluation budget, i.e., the same number of circuit executions (function evaluations). This allows us to isolate the effect of the depth-training strategy from the effect of simply using more optimization calls.

\subsection{Ansatz Layouts}\label{sec:ansatz_layouts}
The ansatz layout plays a central role in VQA performance because it determines the circuit's expressivity, entangling structure \cite{sim2019expressibility}, trainability, and compatibility with the target quantum hardware.
For near-term devices, the hardware compatibility of an ansatz is especially important: even if a circuit is shallow at the logical level, transpilation can substantially increase its depth when the requested two-qubit interactions do not match the native connectivity of the device.

In this work, we target IBM heavy-hex-style connectivity \cite{chamberland2020topological}, which is used by IBM superconducting quantum processors, including Heron-generation devices. The heavy-hex architecture provides sparse nearest-neighbor connectivity; therefore, two-qubit gates between non-adjacent physical qubits must either be avoided by a suitable qubit assignment or implemented through additional routing operations inserted by the transpiler.
Such routing typically introduces SWAP gates, and each SWAP can be decomposed into three two-qubit gates.
Consequently, ansatz layouts that align better with the hardware graph can reduce the transpiled two-qubit gate count and circuit depth, which is important for limiting accumulated noise and decoherence.

We evaluate four ansatz layouts. The first two are hardware-efficient baselines used in previous VQA experiments: the bilinear layout and the colored layout. We also introduce two problem- and hardware-aware layouts: \emph{DGC}, where the entanglement pattern is informed by the portfolio data, and \emph{HNDC}, which is designed to preserve a structured deep-chain interaction pattern while remaining compatible with the hardware connectivity.
The following subsections describe each layout in more detail.

\subsubsection{Bilinear Layout}
\label{sec:bilinear_layout}

We use the bilinear layout as one of the hardware-efficient baseline layouts, following its use in prior VQA studies for dynamic portfolio optimization \cite{Nodar2025ScalingVQE}. The layout arranges the \(N_q\) qubits along a one-dimensional path and applies nearest-neighbor two-qubit gates in two alternating
sublayers. Let the qubits be indexed as \(q_1,\dots,q_{N_q}\). In the first sublayer, two-qubit gates are
applied to the disjoint pairs
\[
(q_1,q_2),\;(q_3,q_4),\;(q_5,q_6),\dots,
\]
and in the second sublayer they are applied to
\[
(q_2,q_3),\;(q_4,q_5),\;(q_6,q_7),\dots.
\]
Since all gates within each sublayer act on disjoint pairs, they can be executed in parallel. Thus, each bilinear ansatz repetition has a two-qubit depth of \(2\) before transpilation. This layout provides a simple hardware-efficient nearest-neighbor entanglement pattern and is illustrated in Figure~\ref{fig:bilinear_circuit}.

\subsubsection{Colored Layout}
\label{sec:colored_layout}

\begin{figure}[t]
    \centering

    \begin{subfigure}[t]{0.18\columnwidth}
        \centering
        \includegraphics[
            width=\linewidth,
            trim={0cm 0cm 0cm 0cm},
            clip
        ]{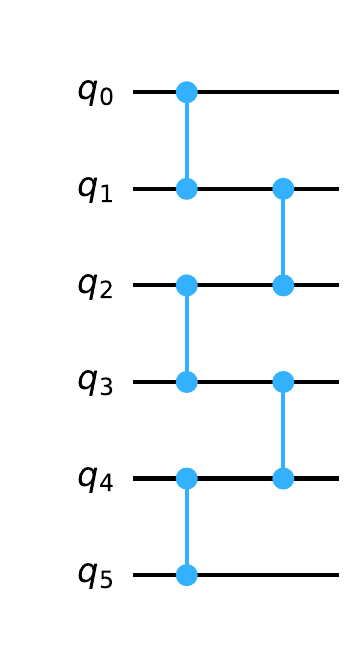}
        \caption{Bilinear layout.}
        \label{fig:bilinear_circuit}
    \end{subfigure}
    \hspace{2cm}
    \begin{subfigure}[t]{0.25\columnwidth}
        \centering
        \includegraphics[
            width=\linewidth,
        ]{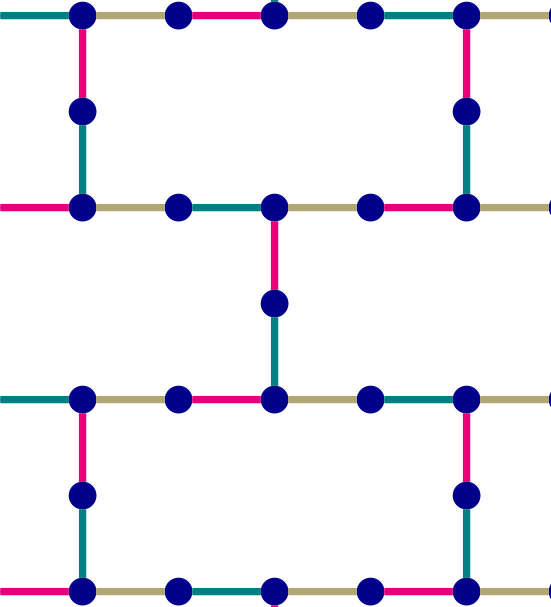}
        \caption{Three-color heavy-hex layout.}
        \label{fig:colored_layout}
    \end{subfigure}

    \caption{
    Baseline hardware-efficient layouts used in this work.
    The bilinear layout applies CZ gates along a one-dimensional nearest-neighbor path using alternating
    sublayers. The colored layout partitions the hardware coupling graph into three edge-color classes,
    where each color class forms a matching and can be executed in parallel.
    }
    \label{fig:baseline_layouts}
\end{figure}

The colored layout used in \cite{kim2023evidence} is a hardware-aware entanglement pattern based on an edge coloring of the target hardware connectivity graph. We use it as a baseline layout following the prior work \cite{Nodar2025ScalingVQE}. Let \(G=(V,E)\) denote the coupling graph of the selected qubits. The edge set \(E\) is partitioned into three disjoint subsets,
\[
E = E_1 \cup E_2 \cup E_3,
\]
such that no two edges within the same subset share a common vertex. Equivalently, this corresponds to a proper 3-edge-coloring of the coupling graph. Each subset \(E_c\) defines one entangling sublayer: two-qubit gates are applied in parallel on all edges in \(E_c\). Since the three color classes are applied sequentially, one colored entangling repetition has two-qubit depth three before transpilation. This layout is illustrated in Figure~\ref{fig:colored_layout}.

\subsubsection{Data-Guided Colored Layout (DGC)}
\label{sec:dgc_layout}

The DGC uses the same three-color entangling schedule as the colored layout, but changes the assignment of problem variables to circuit qubits. Instead of assigning variables in their natural order, DGC chooses the variable-to-qubit mapping so that strongly dependent portfolio variables are more likely to be placed on qubits connected by an entangling edge. Thus, DGC preserves the hardware
efficiency and two-qubit depth of the colored layout while adding a problem-dependent inductive bias.

Let \(f:\mathcal{P}\rightarrow \{1,\dots,N_q\}\) be a bijection from the selected qubits \(\mathcal{P}\) to the \(N_q\) logical variables, where \(f(p)\) denotes the logical variable assigned to qubit \(p\). Given the colored edge sets \(\{E_c\}_{c=1}^{3}\) defined above, DGC chooses \(f\) by maximizing the dependency score placed on the executed two-qubit gates:
\[
\max_f
\;
F(f)
=
\sum_{c=1}^{3}
\sum_{(p,q)\in E_c}
A_{f(p),f(q)}.
\]
Here, \(A\) is a data-derived affinity matrix computed from low-energy candidate solutions. In our implementation, \(A\) is based on weighted mutual information between binary variables, with an entropy softening factor to avoid overemphasizing variables that are nearly frozen in the sampled distribution.
The resulting mapping is used only to place logical variables on circuit qubits. Details of the affinity construction and mapping heuristic are given in Appendix~\ref{app:dgc_mapping}.

\subsubsection{Heavy-Hex Native Deep-Chain Layout (HNDC)}
\label{sec:HNDC_layout}

The bilinear and colored layouts described above are hardware-compatible and have low two-qubit depth. This is advantageous for noise reduction, but it can also limit the expressivity and entangling capability of the ansatz. A direct way to increase expressivity is to repeat the same entangling pattern multiple times. However, this also increases the number of variational parameters linearly with the number of repetitions and makes the classical optimization problem more difficult.

Another possibility is to increase the two-qubit interaction depth within a single repetition while keeping the number of variational parameters fixed. Fully connected layouts or long path-like layouts can increase the logical two-qubit depth before transpilation, but they are generally not compatible with sparse
hardware connectivity. As a result, the transpiler may need to insert many SWAP gates, causing the transpiled two-qubit depth to grow substantially. Even a simple bilinear pattern can require additional routing when the number of logical qubits is close to the full capacity of the target device. We report this effect empirically in Section~\ref{sec:qpu_ansatz}.

The goal of HNDC is to increase the two-qubit interaction depth within a single ansatz repetition while using only hardware-native couplings, so that the added entangling structure does not require routing through non-native edges after transpilation.
The selected hardware graph is partitioned into several disjoint connected components. Each component is designed to have a moderate internal two-qubit depth. This has two advantages: first, each component can support a deeper local entangling pattern; second, enough unused hardware edges remain between components so that different components can be connected in a subsequent entangling layer.

\begin{figure*}[t]
    \centering
    \includegraphics[width=0.7\textwidth]{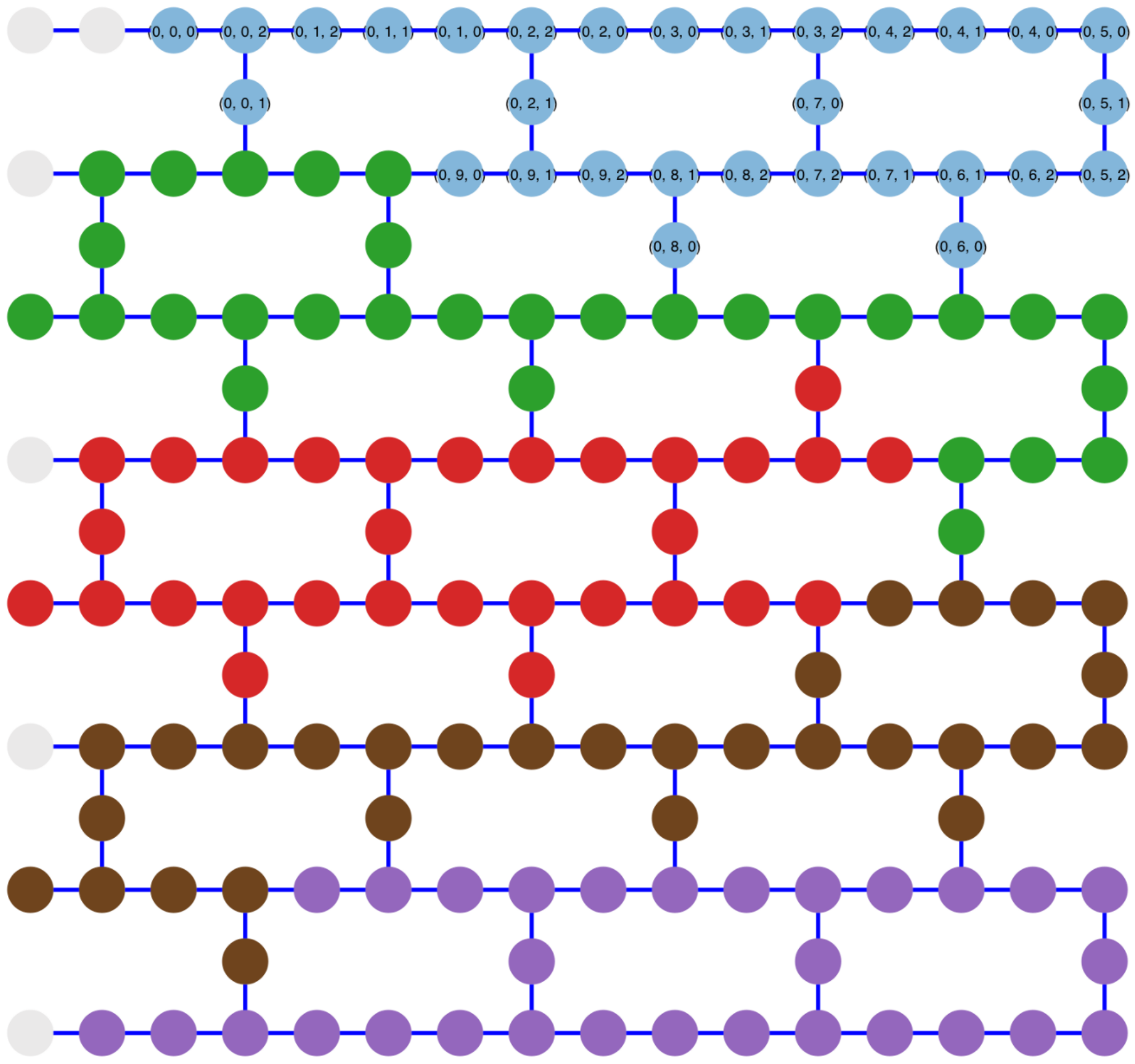}
    \caption{
        Example HNDC mapping. Colors denote different portfolio time steps. The blue component shows the assignment
        for time step \(t=0\), with each qubit labeled by \((t,i,k)\), denoting time index, asset index, and
        binary-encoding bit. $k=2$ is the MSB. The layout increases local two-qubit interaction depth using native hardware edges while preserving
        hardware-compatible inter-component connections.
        }
    \label{fig:hndc_layout}
\end{figure*}

Concretely, the first part of the HNDC repetition applies parallel entangling subcircuits on the disjoint components. The second part uses available hardware edges between components to couple the components together. Thus, the layout forms a connected entangling graph over the selected qubits without relying on non-native long-range interactions. Figure~\ref{fig:hndc_layout} illustrates an example of this construction.

We also incorporate coarse information from the portfolio formulation when assigning logical variables to the HNDC structure. In our objective, the risk term couples assets within the same time step through the period-wise covariance matrix. Therefore, binary variables associated with assets from the same time step are placed preferentially within the same component. In addition, each integer holding is represented using \(K\) binary variables. Since more significant bits have larger numerical weights in the integer encoding, interactions involving higher-order bits can have larger influence on the objective. We therefore place more significant bits on higher-degree nodes when possible, and arrange the bits belonging to the same integer holding close to each other, for example along a short path. An example of this placement rule is shown in the blue component of Figure~\ref{fig:hndc_layout}.

This variable placement is intentionally heuristic. We use only simple local adjustments rather than solving a full placement optimization problem, so that the layout remains easy to adapt to different numbers of assets, time steps, and encoding bits. The purpose is to introduce a problem-aware bias while preserving the main hardware-compatible structure of the layout.

In our experiments, each time step contains \(30\) binary variables, and the HNDC components are chosen so that the logical two-qubit depth within a component is \(30\). Because the entangling edges are selected from the native hardware graph, no additional routing overhead is observed for HNDC in the transpilation settings used in our experiments: the transpiled two-qubit depth matches the logical two-qubit depth. Thus, HNDC increases the number of sequential hardware-native two-qubit gate layers within a single ansatz repetition without adding routing-induced two-qubit depth. We compare HNDC against shallower layouts and other layouts with similar transpiled two-qubit depth in Section~\ref{sec:qpu_ansatz}.

\section{Experiments and Results}
\label{sec:experiments}

This section evaluates the proposed components in two stages. First, we perform three simulator-based studies to select the CVaR schedule, classical optimizer, and ansatz depth strategy. Each simulator study uses the best-performing configuration identified in the preceding study as the baseline for the next comparison. This sequential protocol reduces the experimental search space, but it does not exhaustively test interactions among CVaR scheduling, optimizer choice, and ansatz depth. Second, using the selected configuration, we compare the four ansatz layouts on the \texttt{ibm\_quebec} QPU. 

\subsection{Experimental Protocol}
\label{sec:setup}

\paragraph{Problem instance and objective parameters}
All experiments use a dynamic portfolio instance with \(n=10\) assets, \(m=5\) rebalancing periods, and \(K=3\) binary variables per asset-time holding, giving
\[
N_q = n m K = 150
\]
binary variables, which correspond to the logical qubits used by the sampling-based VQA. The empirical dataset is constructed from daily adjusted closing prices of the ten largest S\&P~500 constituents by market capitalization at the time of dataset construction, obtained using the Yahoo Finance data provider in Qiskit Finance \cite{qiskit_finance_2024} from January 1, 2025 to January 1, 2026. Let \(P\in\mathbb{R}^{n\times T}\) denote the resulting price matrix after preprocessing, where column \(a\) contains the asset prices on trading day \(a\). Returns and covariance matrices are estimated with a rolling window of \(L_{\mathrm{hist}}=60\) trading days. Specifically, we choose \(m+1\) approximately evenly spaced anchor indices \(a_0,\ldots,a_m\) between \(L_{\mathrm{hist}}\) and the last available trading day, set the anchor prices to \(p_t=P_{:,a_t}\), and estimate the period-\(t\) covariance matrix \(\Sigma_t\) from the return series computed over the price window \(P_{:,a_t-L_{\mathrm{hist}}:a_t}\), following the procedure in \cite{buonaiuto2023best, haghighi2025eaqga}. The objective parameters are fixed across all experiments and are chosen to keep the return, risk, transaction-cost, cash, and penalty terms on comparable numerical scales while remaining financially interpretable; the full set is given in Table~\ref{tab:problem_params}.

\begin{table}[t]
\centering
\caption{Problem instance and objective parameters used in the experiments.}
\label{tab:problem_params}
\begin{tabular}{l c l}
\hline
Parameter & Value & Description \\
\hline
\(n\) & 10 & Number of assets \\
\(m\) & 5 & Number of rebalancing periods \\
\(K\) & 3 & Bits per integer holding \\
\(N_q=n m K\) & 150 & Binary variables / qubits \\
\(B\) & 4 & Target active assets per period \\
\(h_{\max}\) & 5 & Maximum allowed holding \\
\(L_{\mathrm{hist}}\) & 60 & Historical estimation window \\
\(\lambda\) & \(10^{-1}\) & Risk-aversion coefficient \\
\(\delta\) & \(10^{-2}\) & Transaction cost coefficient \\
\(\nu\) & \(10^{-3}\) & Cash interest rate \\
\(W\) & \(10^{4}\) & Initial cash budget \\
\(\alpha_{\mathrm{card}}\) & 10 & Cardinality penalty multiplier \\
\(\alpha_{\mathrm{hold}}\) & 10 & Holding-cap penalty multiplier \\
\hline
\end{tabular}
\end{table}

\paragraph{Penalty calibration}
The model uses soft penalties for matching the target cardinality \(B\) and discouraging holdings above \(h_{\max}\). To avoid manual tuning, penalty coefficients are scaled using a conservative objective scale \(S_{\mathrm{obj}}\). Specifically, we set \(P_{\mathrm{card}}=\alpha_{\mathrm{card}}S_{\mathrm{obj}}\) and
\(P_{\mathrm{hold}}=\alpha_{\mathrm{hold}}S_{\mathrm{obj}}\), using fixed multipliers shared across all experiments. Details are given in Appendix~\ref{app:penalty_cal}.

\paragraph{Classical reference}
As a classical reference, we solve the implemented mixed-integer optimization model using SCIP \cite{SCIPOptSuite10}. For the fixed instance studied in this work, SCIP terminates by matching primal and dual bounds at $-1456.811$. We therefore use this value as the certified optimum of the implemented classical model for this instance. SCIP is an open-source solver for mixed-integer and constraint integer optimization, based on a branch-and-cut framework with presolving, cutting planes, primal heuristics, and constraint propagation. The red dashed line in the result figures denotes this SCIP reference value.

\paragraph{Circuit, sampling, and budget} \label{sec:shot_scaling}
Across both simulator and QPU experiments, the variational circuit uses a TwoLocal-style gate structure consisting of layers of parameterized \(R_y\) rotations followed by CZ entangling gates. The placement of the CZ gates is
determined by the ansatz layout used in each experiment. Unless otherwise stated, all variational parameters are initialized to \(\pi/4\) and constrained to the periodic interval \([0,2\pi]\) during optimization.

One objective evaluation consists of sampling the parameterized circuit, decoding measured bitstrings into integer holdings, evaluating the portfolio objective on each decoded sample, and aggregating the resulting costs using CVaR. In the plots and tables, we refer to one objective-function evaluation as one \emph{iteration}. Each configuration is run with a total budget of \(1000\) iterations.

Since CVaR at level \(\alpha\) uses only the lowest-cost \(\alpha\)-fraction of samples, we scale the shot count as
\[
N_{\mathrm{shots}}(\alpha)=\left\lceil \frac{N_0}{\alpha}\right\rceil,
\]
with base shot count \(N_0=2000\), motivated by prior work on CVaR-based estimates from noisy samples \cite{Barron2024ProvableBounds}.

\paragraph{Postprocessing}
For metrics reported as postprocessed, we apply a lightweight bit-flip postprocessing procedure similar to the one used in \cite{Agliardi2025SamplingVQA}. For each sampled bitstring, the maximum number of attempted bit flips is capped at the number of binary variables. Bits are visited in random order; if a bit flip improves the objective value, the move is accepted and the search restarts with a new random ordering. The procedure terminates when no improving flip is found within the flip budget.

\paragraph{Metrics}
For simulator studies, the primary metric is the median, across independent runs, of the best objective value observed at the final iteration. As secondary metrics, we report the CVaR objective trajectory and the final CVaR value. For QPU experiments, we report the corresponding final-circuit metrics for each layout, including the final CVaR value, the best raw sampled objective value, the best postprocessed objective value, and KDEs of raw and postprocessed sampled costs.

\subsection{Simulator Studies}
\label{sec:simulator_studies}

The first three studies are controlled simulator-based ablations. We use a Matrix Product State (MPS)
simulator \cite{qiskit_mps_2024, vidal2003efficient} and a fixed bilinear layout in Studies I--III, so
that each study isolates one design choice: CVaR scheduling, optimizer scheduling, or ansatz depth strategy.
Each simulator configuration is repeated over \(10\) independent runs. The best-performing configuration
from each study is carried forward to the next study.

\subsubsection{Study I: Adaptive CVaR vs Constant CVaR Baselines}
\label{sec:study_cvar}

This study evaluates the effect of the CVaR aggregation strategy. We compare fixed CVaR levels
\(\alpha\in\{0.1,0.2,0.3,1.0\}\), the proposed adaptive CVaR schedule, and an additional
\(\alpha=0.1\) configuration without shot scaling. All configurations use the same ansatz layout, optimizer,
objective, and total evaluation budget. NFT is used as the optimizer in this study, and for the adaptive
CVaR schedule the CVaR level is kept fixed during the evaluations associated with each coordinate-wise NFT
update.

For the adaptive CVaR schedule, we set
\(\alpha_{\max}=1.0\), \(\alpha_{\min}=0.1\), and decrease \(\alpha\) by \(\Delta\alpha=0.1\) every 48 NFT parameter updates, until it reaches \(\alpha_{\min}=0.1\), where it remains fixed.

\begin{figure*}[t]
    \centering

    \begin{subfigure}[t]{0.32\textwidth}
        \centering
        \includegraphics[width=\linewidth]{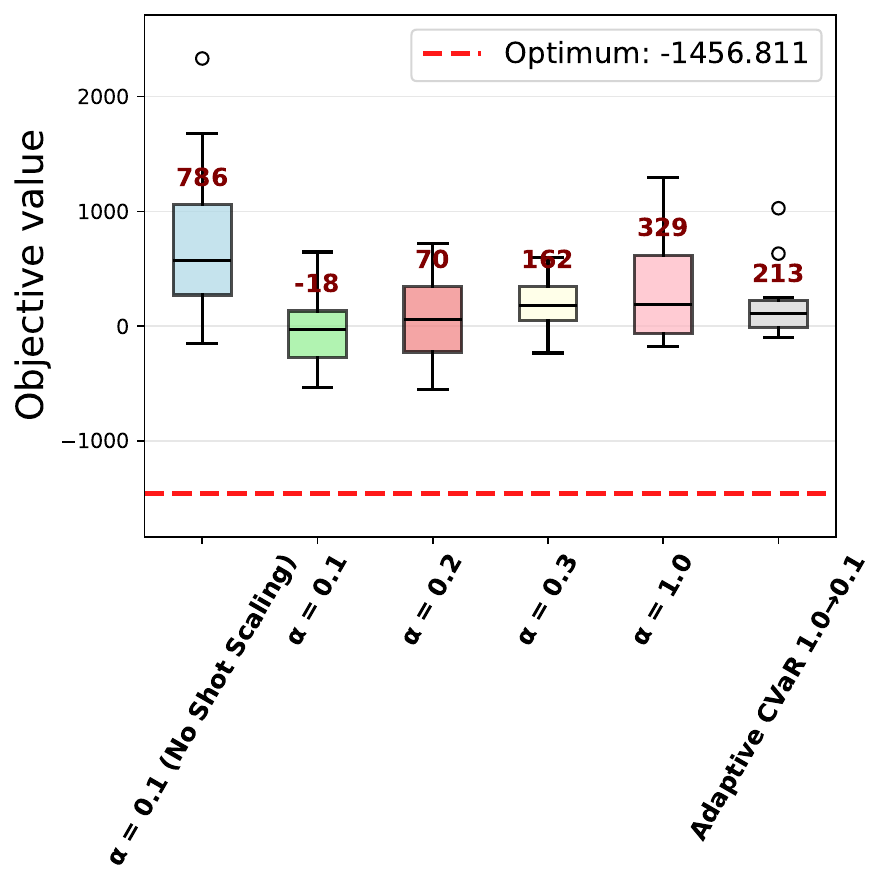}
        \caption{}
        \label{fig:cvar_final_cost}
    \end{subfigure}
    \hfill
    \begin{subfigure}[t]{0.31\textwidth}
        \centering
        \includegraphics[width=\linewidth]{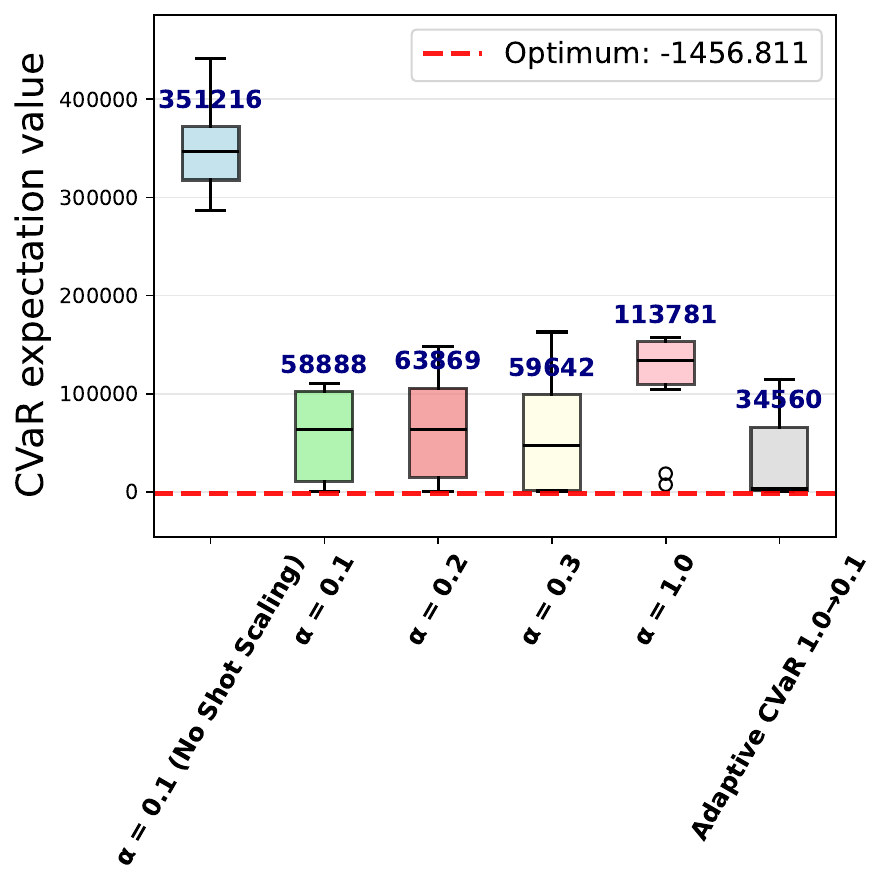}
        \caption{}
        \label{fig:cvar_final_cvar}
    \end{subfigure}
    \hfill
    \begin{subfigure}[t]{0.29\textwidth}
        \centering
        \raisebox{15mm}{%
            \includegraphics[width=\linewidth]{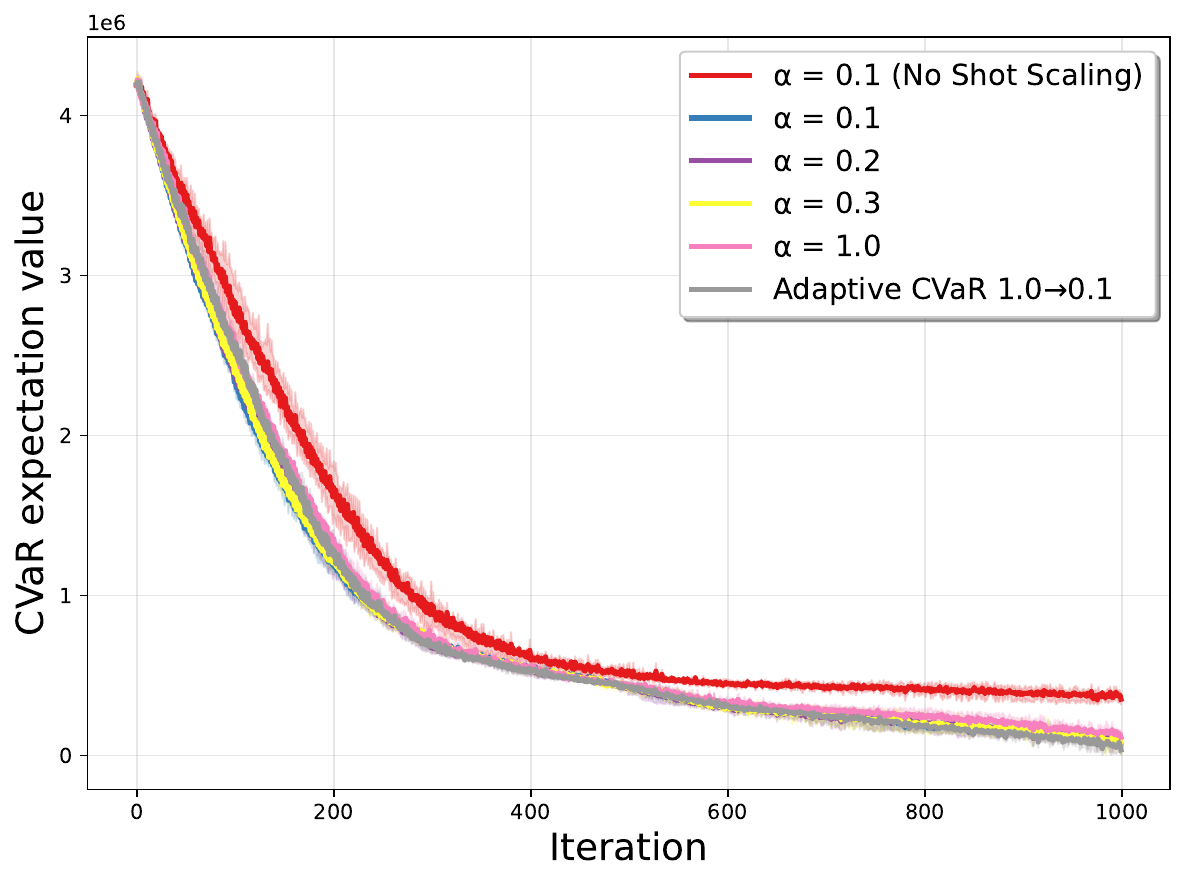}%
        }
        \caption{}
        \label{fig:cvar_cost_evolution}
    \end{subfigure}

    \caption{
    Comparison of fixed and adaptive CVaR strategies over 10 independent runs.
    (a) Best objective value observed at the final iteration.
    (b) Final CVaR value evaluated using a common \(\alpha=0.1\) level for all configurations.
    In (a) and (b), the horizontal line inside each box indicates the median; the annotated value is the mean across runs.
    (c) Evolution of the mean CVaR objective evaluated using the same common \(\alpha=0.1\) level.
    Lower values indicate better portfolio objective values.
    }
    \label{fig:cvar_results}
\end{figure*}

Figure~\ref{fig:cvar_results} summarizes the results. Among fixed-CVaR baselines with shot scaling, smaller CVaR levels generally improve the median final best objective value, with \(\alpha=0.1\) giving the lowest median value. The adaptive schedule remains competitive, but does not achieve the best median value under
this best-solution metric. The comparison between \(\alpha=0.1\) with and without shot scaling shows a substantial degradation when shot scaling is removed, indicating that low-CVaR optimization requires enough samples to estimate the low-cost tail reliably.

When all final circuits are evaluated using a common \(\alpha=0.1\) CVaR level, the adaptive CVaR schedule achieves the best final CVaR value. This indicates that the adaptive schedule improves the quality of the low-cost tail of the sampled distribution, even though fixed \(\alpha=0.1\) finds the best individual
solutions more effectively. The adaptive schedule also uses larger CVaR levels during earlier iterations and therefore requires fewer shots on average than a fixed \(\alpha=0.1\) strategy. The CVaR evolution in Figure~\ref{fig:cvar_cost_evolution} decreases overall, with fluctuations expected from NFT's shifted parameter evaluations during coordinate-wise updates.

\subsubsection{Study II: Optimizer Scheduling}
\label{sec:study_optimizer}

This study evaluates the effect of the classical optimizer. We compare NFT, PSO, COBYLA, and the proposed PSO\(\rightarrow\)NFT two-stage optimizer. Based on Study~I, all configurations use fixed CVaR with \(\alpha=0.1\).

For the PSO\(\rightarrow\)NFT optimizer, the PSO phase is allocated a minimum budget of \(180\) objective-function evaluations and a maximum budget of \(600\) evaluations, corresponding to \(30\%\) of the maximum PSO budget and \(60\%\) of the total \(1000\)-evaluation budget, respectively. After the minimum PSO budget is reached, stagnation is checked using a sliding window of \(100\) evaluations. If the relative improvement in the best objective value over this window falls below \(10\%\), PSO is terminated early and the remaining evaluation budget is transferred to NFT. PSO is configured with \(20\) particles, inertia weight \(w=0.7\), and cognitive and social coefficients \(c_1=c_2=1.4\). Particle positions are constrained to \([0,2\pi]\) in each parameter dimension, and the maximum particle velocity is capped at \(20\%\) of the parameter span.

\begin{figure*}[t]
    \centering

    \begin{subfigure}[t]{0.30\textwidth}
        \centering
        \includegraphics[width=\linewidth]{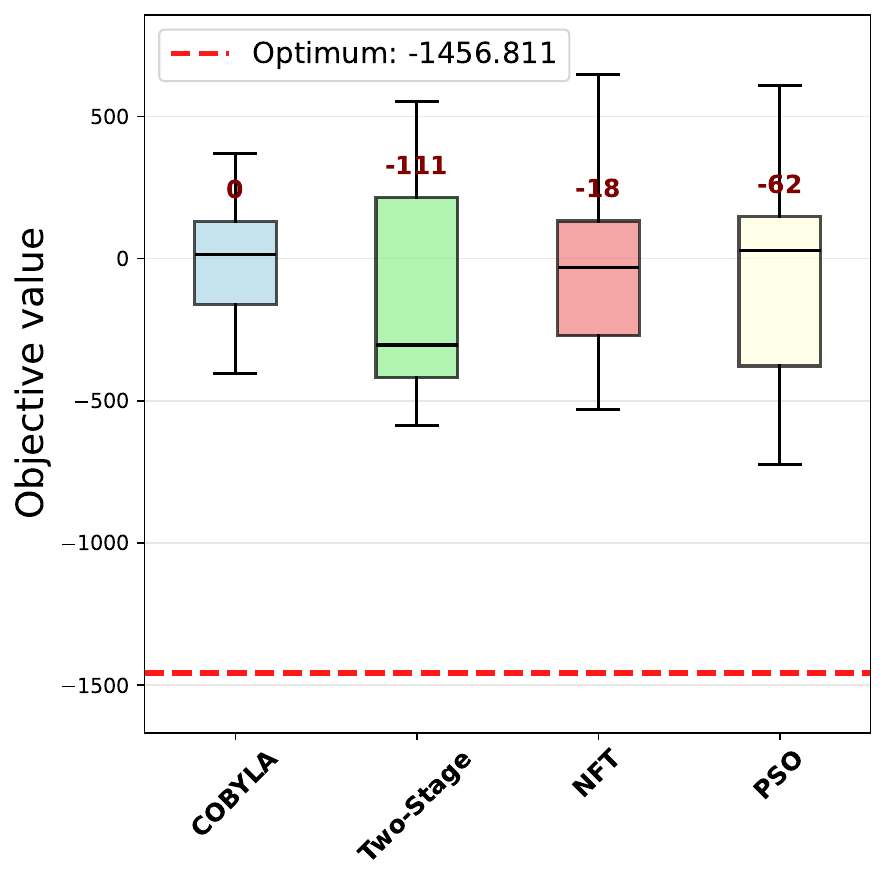}
        \caption{}
        \label{fig:optimizer_final_cost}
    \end{subfigure}
    \hfill
    \begin{subfigure}[t]{0.30\textwidth}
        \centering
        \includegraphics[width=\linewidth]{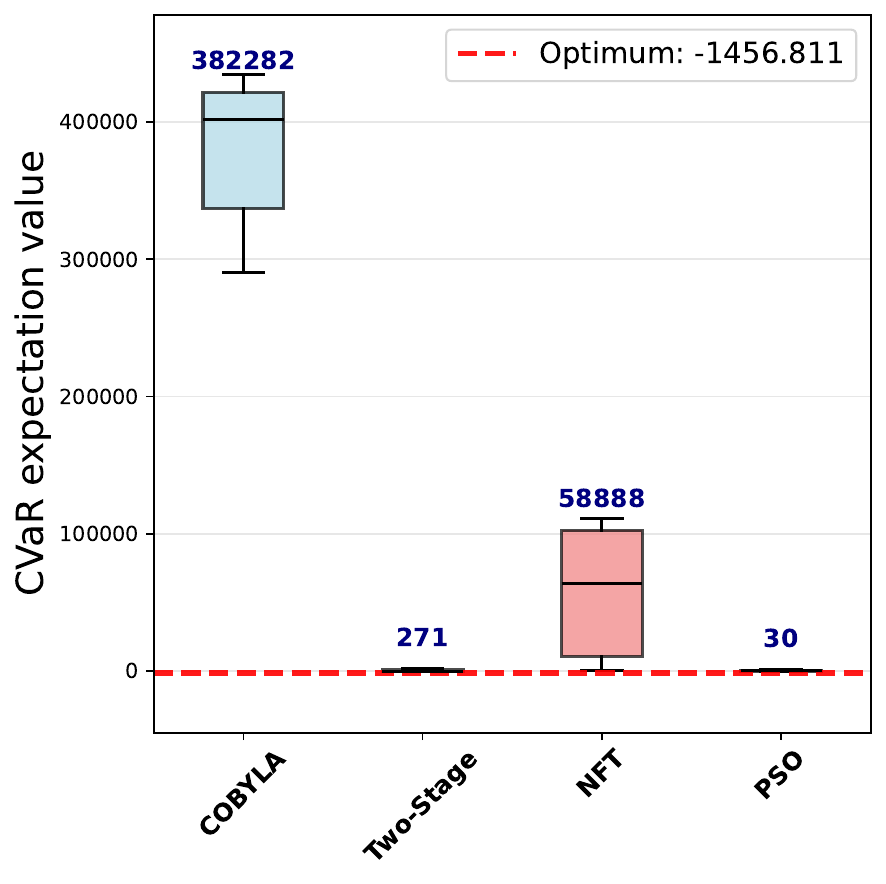}
        \caption{}
        \label{fig:optimizer_final_cvar}
    \end{subfigure}
    \hfill
    \begin{subfigure}[t]{0.32\textwidth}
        \centering
        \raisebox{4mm}{%
            \includegraphics[width=\linewidth]{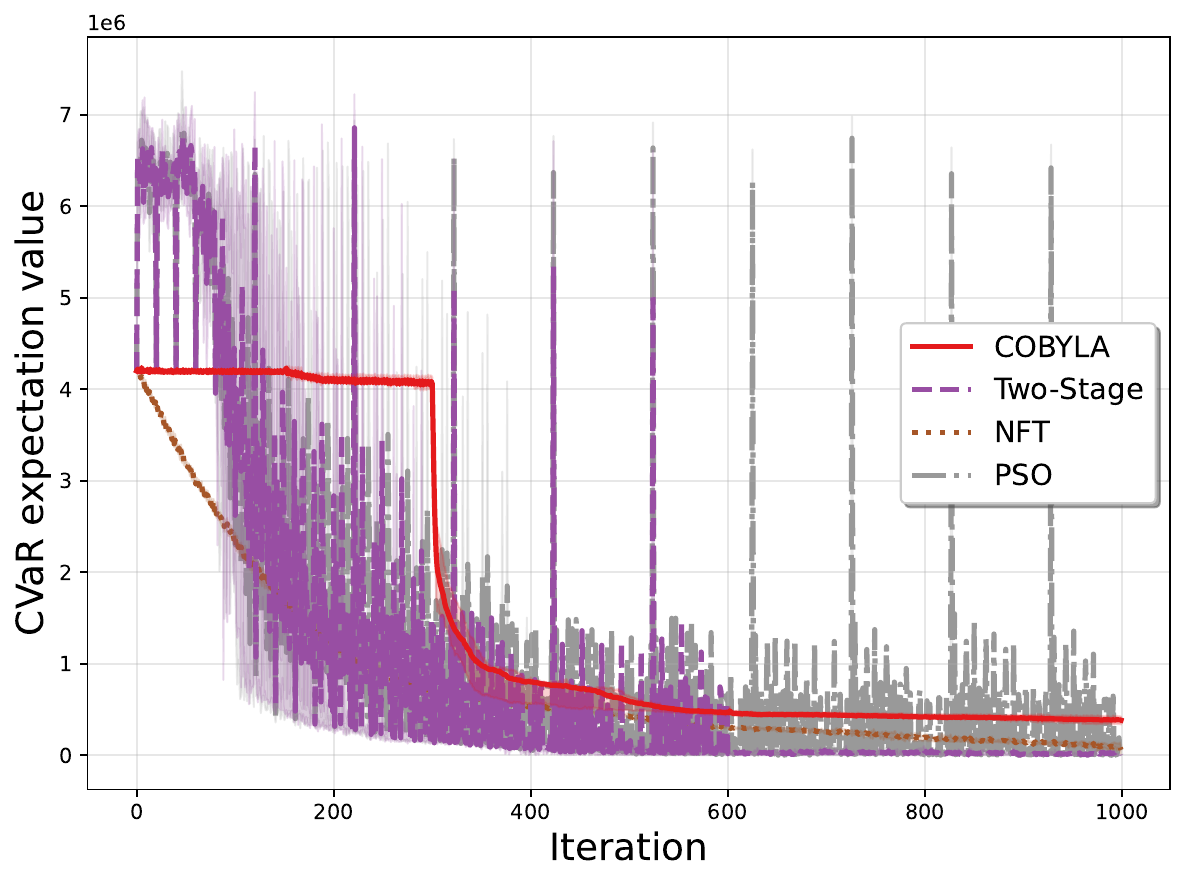}%
        }
        \caption{}
        \label{fig:optimizer_evolution}
    \end{subfigure}

    \caption{
    Comparison of classical optimizers under fixed CVaR level \(\alpha=0.1\) over 10 independent runs.
    (a) Best objective value observed at the final iteration.
    (b) Final CVaR value of the sampled distribution.
    In (a) and (b), the horizontal line inside each box indicates the median; the annotated value is the mean across runs.
    (c) Evolution of the mean CVaR objective during optimization.
    Lower values indicate better portfolio objective values.
    }
    \label{fig:optimizer_results}
\end{figure*}

The results are shown in Figure~\ref{fig:optimizer_results}. The PSO\(\rightarrow\)NFT optimizer achieves the
lowest median final best objective value, followed by NFT, COBYLA, and PSO. These rankings are based on the medians; the annotated means may differ because of outlier runs. In the final CVaR metric, PSO is slightly better than PSO\(\rightarrow\)NFT, while both PSO-based methods outperform NFT and COBYLA. We therefore select PSO\(\rightarrow\)NFT for subsequent experiments because the primary selection criterion is the final best objective value, while noting that PSO and PSO\(\rightarrow\)NFT are close overall.

The convergence trajectories in Figure~\ref{fig:optimizer_evolution} reflect the expected behavior of the optimizers. PSO produces larger fluctuations because the swarm evaluates diverse regions of the parameter space. NFT gives a smoother trajectory, consistent with its coordinate-wise local update rule. The PSO\(\rightarrow\)NFT schedule initially follows the exploratory behavior of PSO and becomes more stable after switching to NFT.

\subsubsection{Study III: Depth and Sequential Growth}
\label{sec:study_depth}

This study evaluates the effect of ansatz repetition depth and sequential growth. Based on the previous studies, we use fixed CVaR with \(\alpha=0.1\) and the PSO\(\rightarrow\)NFT optimizer. We compare fixed-depth ansatzes with one, two, and three repetitions against the sequential-growth strategy described in
Section~\ref{sec:depth_strategy}.

For sequential growth, the first layer is optimized for \(400\) iterations using PSO\(\rightarrow\)NFT. A second layer is then appended and optimized for another \(400\) iterations while keeping the first layer fixed. Finally, all parameters are released and jointly refined for \(200\) additional iterations using NFT.
We use NFT rather than the two-stage optimizer in this final refinement stage because the parameters have already been partially optimized, making local refinement more appropriate.
In addition, the remaining budget is relatively small, leaving limited room for a new global-search phase.

\begin{figure*}[t]
    \centering

    \begin{subfigure}[t]{0.30\textwidth}
        \centering
        \includegraphics[width=\linewidth]{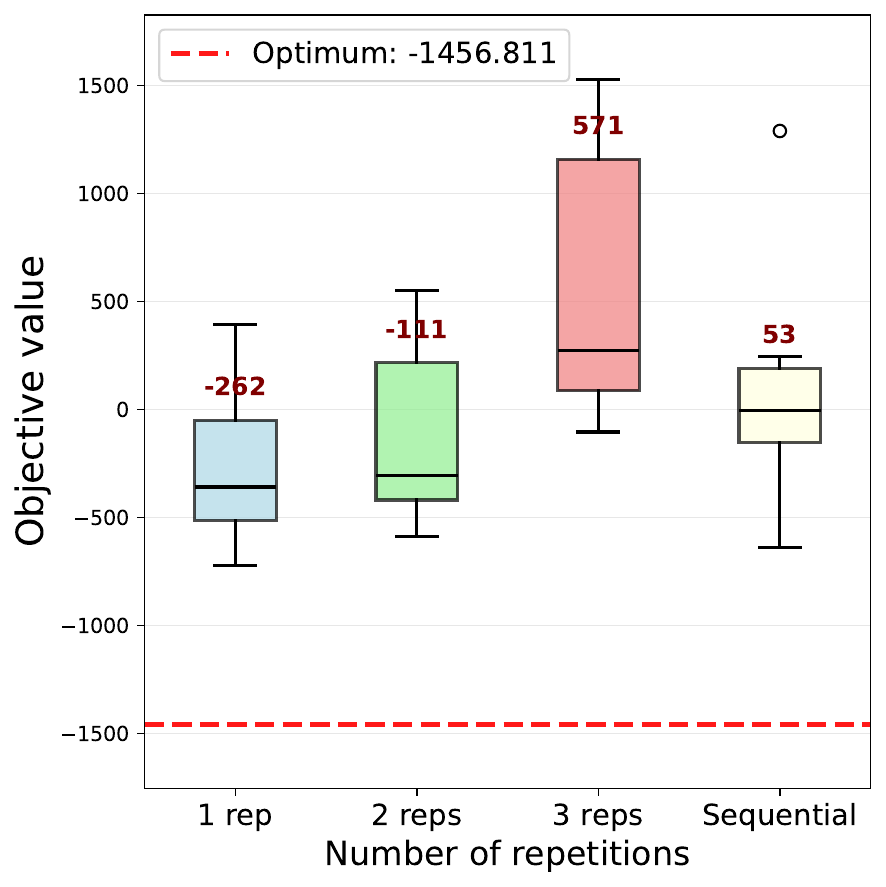}
        \caption{}
        \label{fig:depth_final_cost}
    \end{subfigure}
    \hfill
    \begin{subfigure}[t]{0.30\textwidth}
        \centering
        \includegraphics[width=\linewidth]{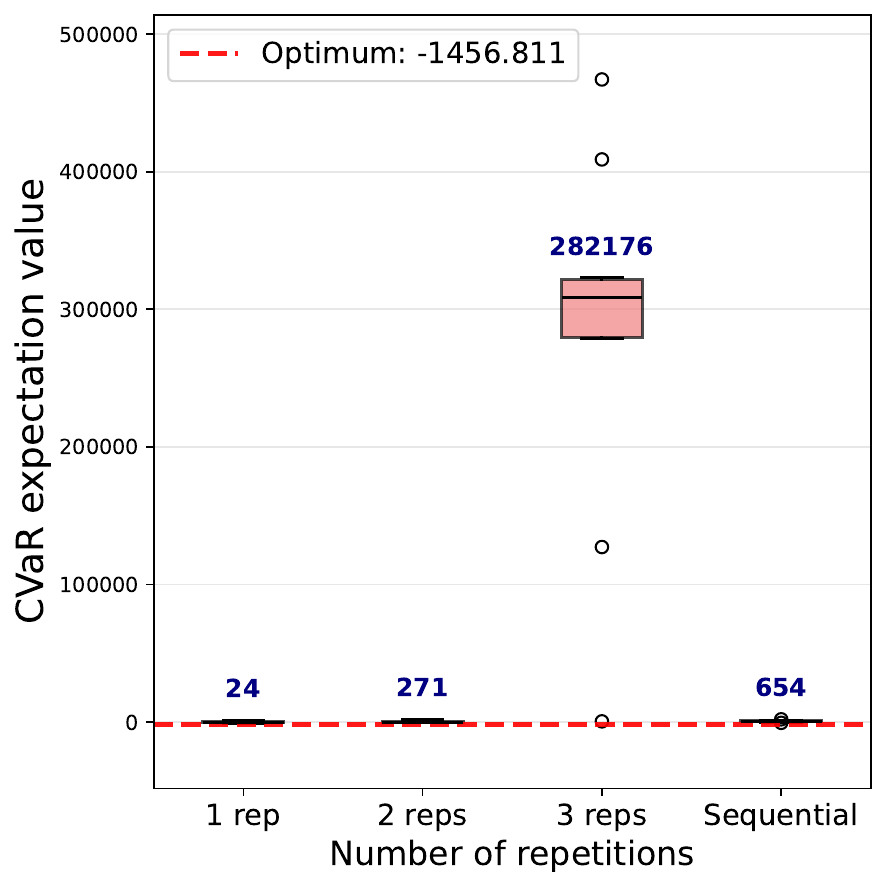}
        \caption{}
        \label{fig:depth_final_cvar}
    \end{subfigure}
    \hfill
    \begin{subfigure}[t]{0.32\textwidth}
        \centering
        \raisebox{2mm}{%
            \includegraphics[width=\linewidth]{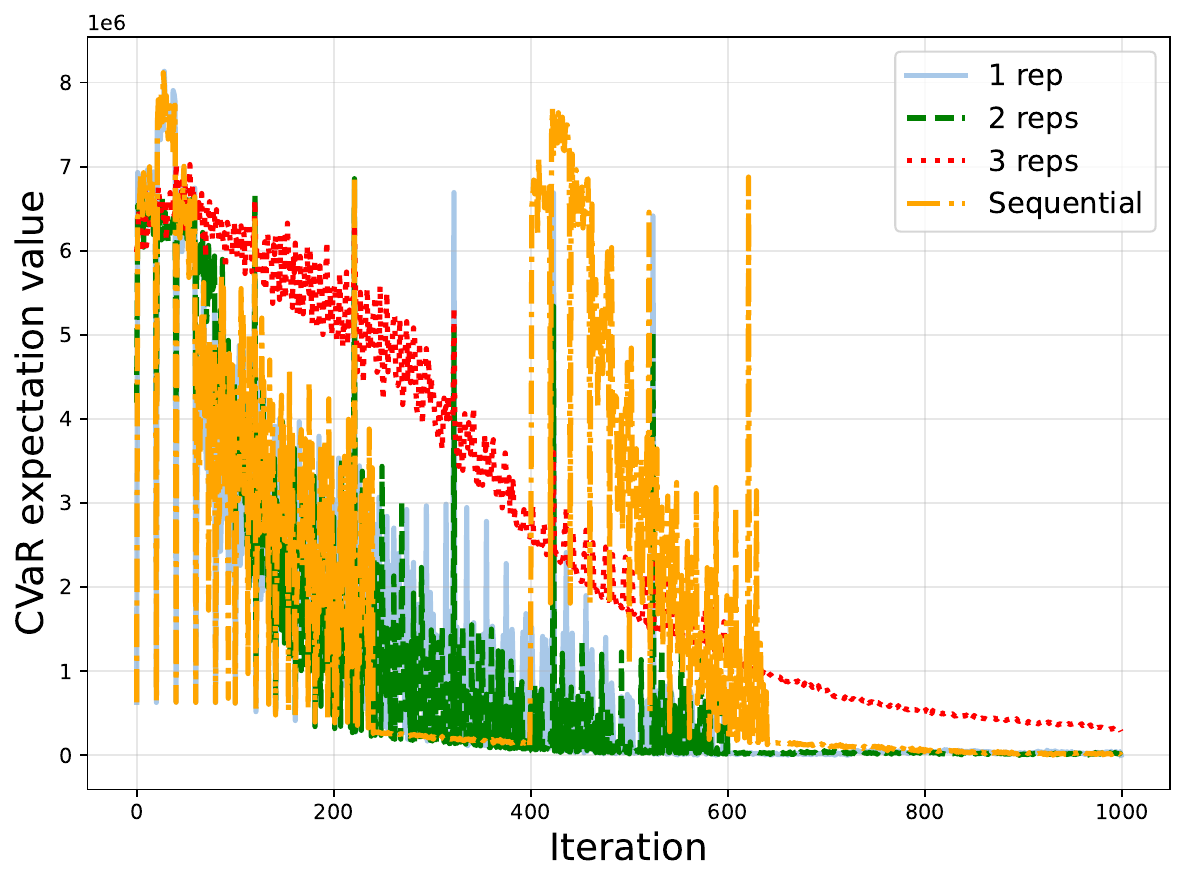}%
        }
        \caption{}
        \label{fig:depth_evolution}
    \end{subfigure}

    \caption{
    Comparison of ansatz repetition depth and sequential growth under fixed CVaR level \(\alpha=0.1\) and
    the PSO\(\rightarrow\)NFT optimizer over 10 independent runs.
    (a) Best objective value observed at the final iteration.
    (b) Final CVaR value of the sampled distribution.
    In (a) and (b), the horizontal line inside each box indicates the median; the annotated value is the mean across runs.
    (c) Evolution of the mean CVaR objective during optimization.
    Lower values indicate better portfolio objective values.
    }
    \label{fig:depth_results}
\end{figure*}

Figure~\ref{fig:depth_results} summarizes the results. The one-repetition ansatz achieves the best median final objective value, with the two-repetition ansatz close behind. The sequential-growth strategy performs worse than the fixed two-repetition ansatz, and the three-repetition ansatz gives the weakest performance. A
similar ordering is observed in the final CVaR metric.

The result suggests that, under the fixed evaluation budget used here, additional repetitions do not necessarily improve performance. One, two, and three repetitions contain \(150\), \(300\), and \(450\)
variational parameters, respectively. With the same number of objective evaluations, ansatzes with fewer repetitions allow more frequent effective updates per parameter under the same evaluation budget, which may explain the advantage of the one-repetition circuit.
The sequential-growth result also suggests that fixing earlier layers can limit co-adaptation between layers, even when a final joint NFT refinement is applied. The CVaR evolution in Figure~\ref{fig:depth_evolution} shows that the one-repetition ansatz starts from a better objective value and maintains its advantage through
most of the optimization.

\subsection{QPU Study: Layout Comparison}
\label{sec:qpu_ansatz}
We perform the layout comparison on \texttt{ibm\_quebec}, an IBM Heron r2
superconducting processor with 156 physical qubits arranged in a heavy-hex connectivity graph. The backend's native basis gates are \{\texttt{cz}, \texttt{id}, \texttt{rx}, \texttt{rz}, \texttt{rzz}, \texttt{sx}, \texttt{x}\}.
Using the best configuration selected from the simulator studies, we compare four ansatz layouts: bilinear, colored, DGC, and HNDC. The hardware experiments use \(\alpha=0.1\),
corresponding to \(N_{\mathrm{shots}}=20000\) shots per objective evaluation under the shot-scaling rule in Section~\ref{sec:shot_scaling}. We also use the PSO\(\rightarrow\)NFT optimizer selected in Study~II. For all QPU executions, dynamic decoupling with an XY4 sequence is applied to reduce idle-qubit errors
\cite{dynamic_decoupling, ibm_error_mitigation_2024}.

Table~\ref{tab:qpu_layout_depths} summarizes the circuit properties before and after transpilation. Colored and DGC preserve their two-qubit depth after transpilation because their entangling edges are defined
directly on the hardware-compatible colored coupling graph; DGC changes only the problem-variable-to-qubit assignment. By contrast, the bilinear layout has two-qubit depth \(2\) before transpilation but depth \(27\) after transpilation, reflecting the routing overhead required to embed a long path near the full device
capacity.

HNDC is designed to use native heavy-hex-compatible edges while increasing the two-qubit interaction depth within a single ansatz repetition. HNDC-1 has two-qubit depth \(30\) before and after transpilation, while HNDC-2 has two-qubit depth \(59\) before and after transpilation. For HNDC-2, the variational parameters are
initialized at \(\pi/8\) rather than \(\pi/4\), so that applying the same entangling structure twice does not start from an overly strong initial rotation configuration.

\begin{table}[t]
\centering
\caption{Circuit properties of the ansatz layouts before and after transpilation on \texttt{ibm\_quebec}. Reported gate counts include measurement operations. Each single-qubit rotation layer contributes 150 parameters per repetition.}
\label{tab:qpu_layout_depths}
\begin{tabular}{l c c c c}
\hline
Layout & Stage & Gates & 2Q gates & 2Q depth \\
\hline
Bilinear & Before & 449 & 149 & 2 \\
Bilinear & After  & 1705 & 378 & 27 \\
\hline
Colored & Before & 470 & 170 & 3 \\
Colored & After  & 920 & 170 & 3 \\
\hline
DGC & Before & 470 & 170 & 3 \\
DGC & After  & 920 & 170 & 3 \\
\hline
HNDC-1 & Before & 470 & 170 & 30 \\
HNDC-1 & After  & 920 & 170 & 30 \\
\hline
HNDC-2 & Before & 790 & 340 & 59 \\
HNDC-2 & After  & 1690 & 340 & 59 \\
\hline
\end{tabular}
\end{table}

\begin{figure*}[t]
    \centering
    \includegraphics[width=1.0\textwidth]{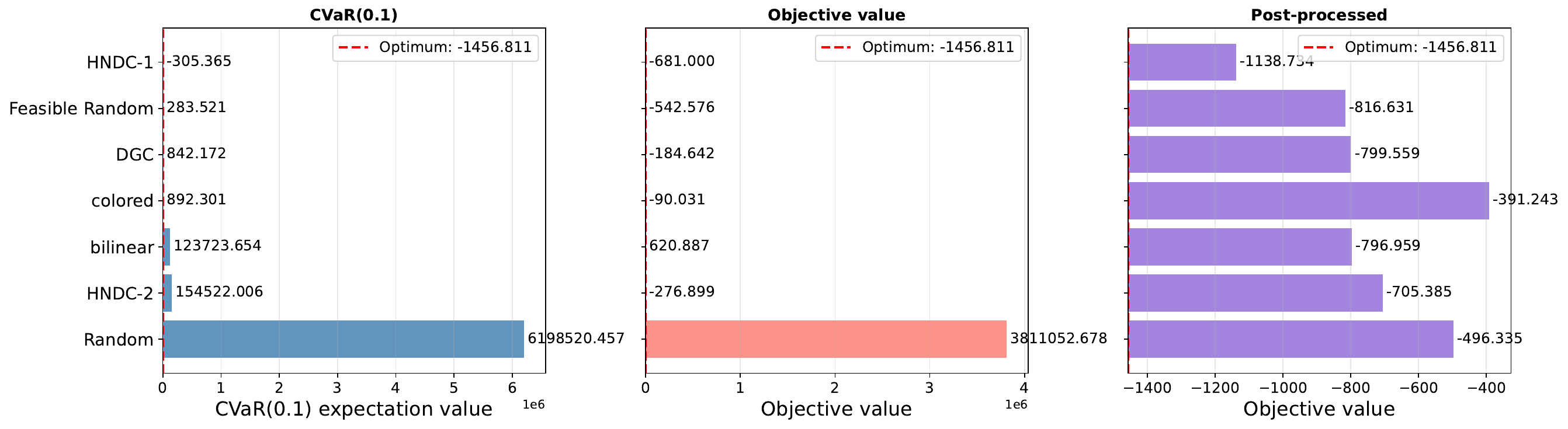}
    \caption{
    QPU performance comparison across ansatz layouts on \texttt{ibm\_quebec}.
    Left: final CVaR value of the sampled distribution.
    Middle: best sampled objective value obtained from the final optimized circuit.
    Right: best objective value after bit-flip postprocessing of the final circuit measurements.
    The figure also includes unrestricted random and feasible-random baselines generated with the same number
    of samples as the final circuit measurements. Lower values indicate better portfolio objective values.
    }
    \label{fig:qpu_summary}
\end{figure*}

In the reported QPU layout comparison (Figure~\ref{fig:qpu_summary}), HNDC-1 achieves the best value among the evaluated layouts across all three reported metrics: final CVaR, best sampled objective, and best postprocessed objective. DGC slightly improves over the standard colored layout across the same metrics, suggesting that data-guided problem-variable-to-qubit assignment can provide a useful inductive bias without increasing two-qubit depth. Bilinear performs worst among the one-repetition layouts in final CVaR and best sampled objective, consistent with its larger post-transpilation routing overhead. HNDC-2 underperforms HNDC-1 and also ranks below DGC in best objective value, consistent with the simulator depth study: additional repetitions may increase expressivity,
but also increase the number of trainable parameters under a fixed evaluation budget.

Figure~\ref{fig:qpu_summary} also includes unrestricted random and feasible-random baselines generated with the same number of samples as the final optimized circuits. All ansatz layouts substantially outperform unrestricted random sampling in final CVaR and best sampled objective value, and all layouts except colored also outperform it after post-processing. More importantly, colored, DGC, and HNDC produce results comparable to feasible-random sampling, despite the fact that feasibility is not explicitly encoded in the
circuit. Since the feasible fraction of the full binary space is approximately \(2.73\times10^{-20}\), this indicates that the optimized circuits are not sampling randomly, but are biased toward regions relevant to the constrained portfolio objective. HNDC-1 also improves over feasible random sampling in final \(\alpha=0.1\) CVaR and best-sample objective value.

\begin{figure}[t]
    \centering
    \includegraphics[width=0.5\columnwidth]{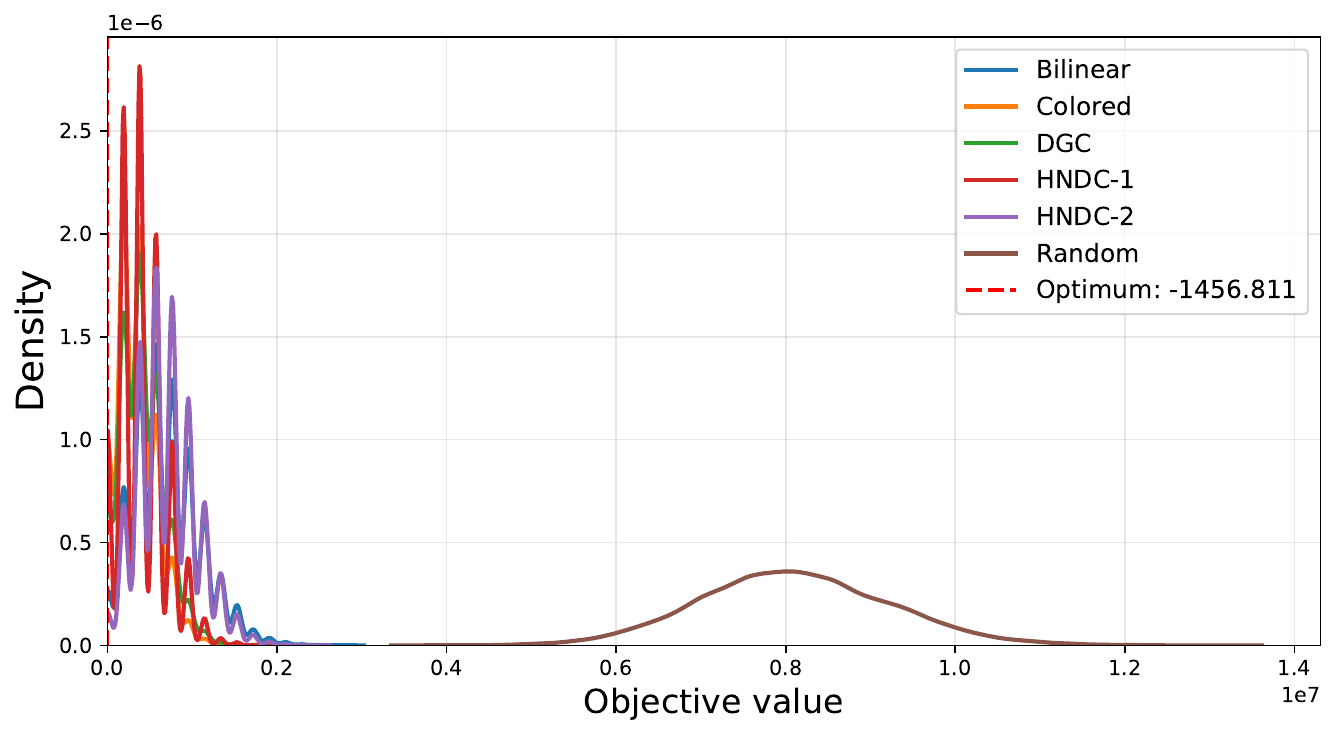}
    \caption{
    Kernel density estimates of raw sampled objective values obtained from the final optimized circuits,
    together with random-sampling baselines. Lower objective values correspond to better portfolio solutions.
    }
    \label{fig:raw_kde}
\end{figure}

The distributional plots provide additional context beyond the final summary metrics.  Figure~\ref{fig:raw_kde} compares raw sampled objective values from the final optimized circuits with unrestricted random sampling. The random samples are far from the optimized distributions, which reflects the difficulty of reaching feasible or near-feasible configurations by sampling uniformly from the full
binary search space. The optimized layouts shift probability mass substantially toward lower objective values relative to this unrestricted random baseline.

\begin{figure}[t]
    \centering
    \includegraphics[width=0.5\columnwidth]{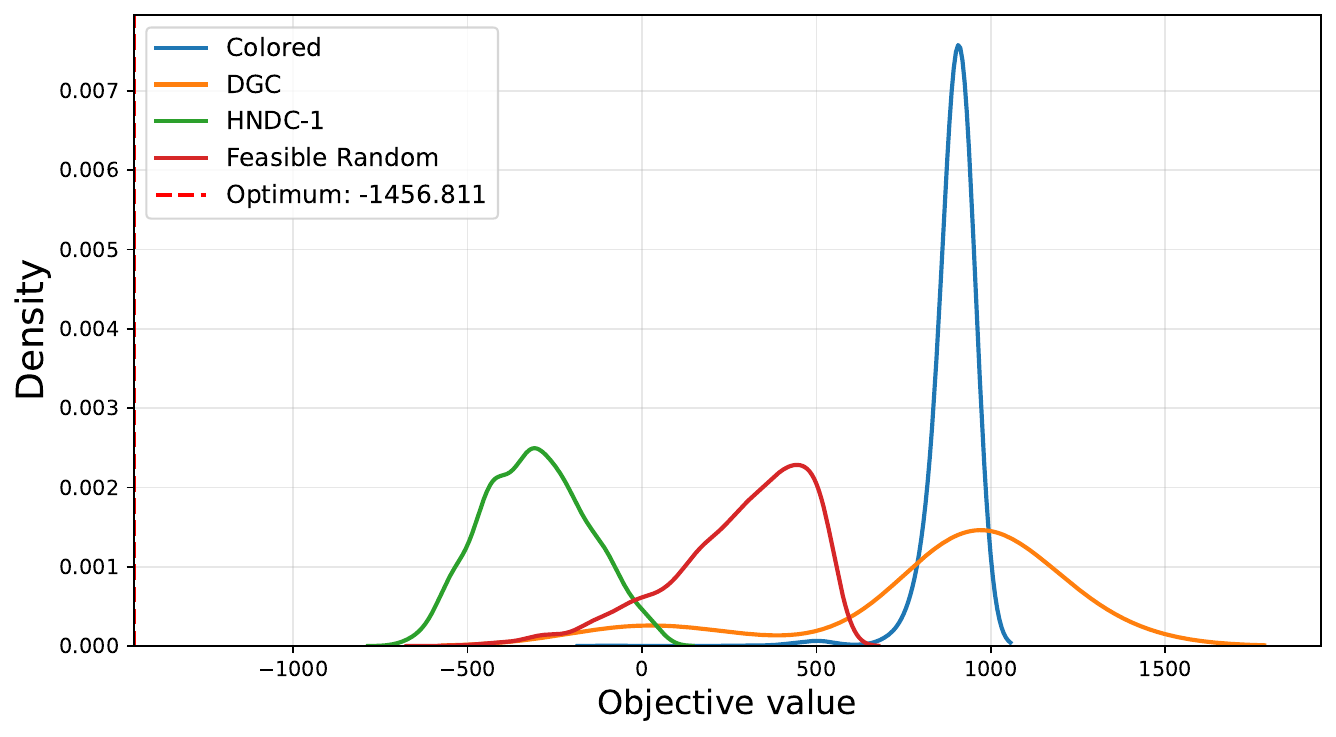}
    \caption{
    Kernel density estimates of raw sampled objective values restricted to the final \(\alpha=0.1\) CVaR tail, with feasible random sampling included as a baseline. Random sampling, Bilinear and HNDC-2 are omitted for readability because their distributions are shifted relative to the main group. Lower values correspond to better portfolio solutions.
    }
    \label{fig:qpu_tail_kde}
\end{figure}

\begin{figure}[t]
    \centering
    \includegraphics[width=0.5\columnwidth]{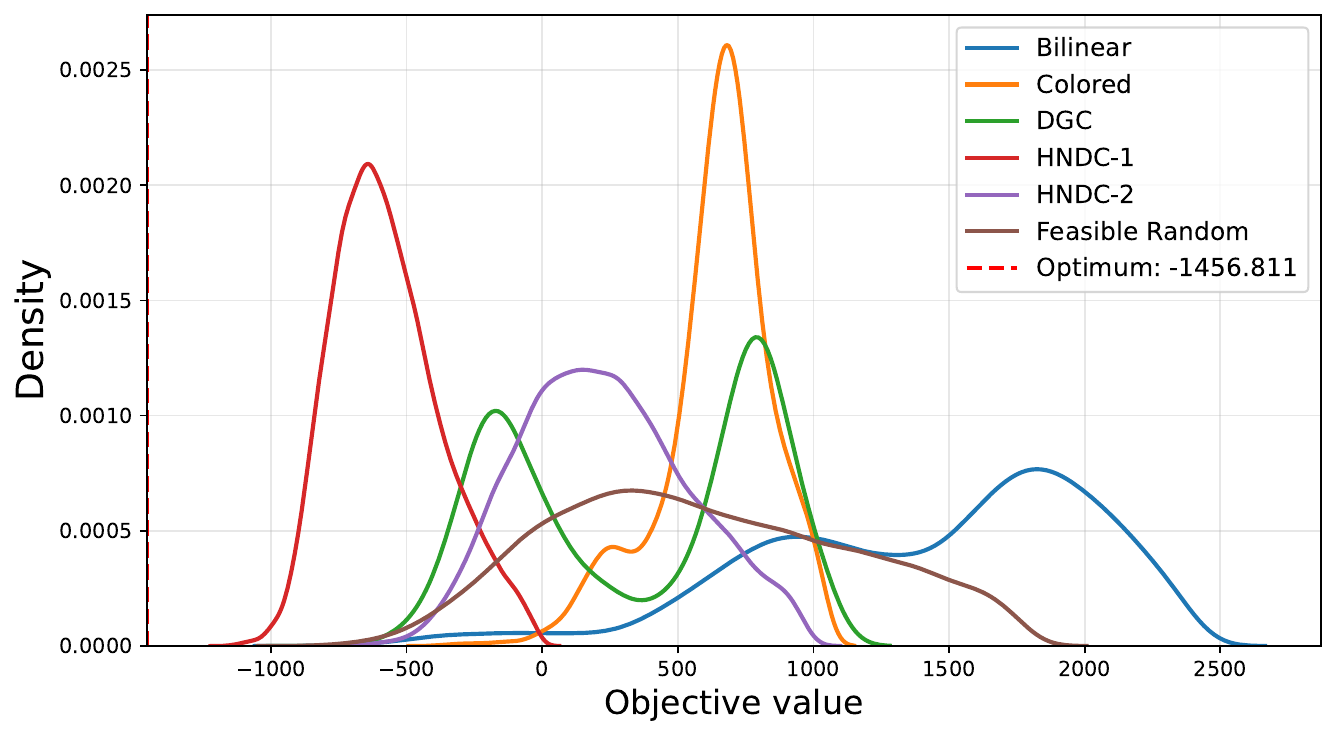}
    \caption{
    Kernel density estimates of postprocessed objective values from the final optimized circuits, with postprocessed feasible random sampling included as a baseline. For visualization, the plot is truncated to the best \(90\%\) of displayed objective values to reduce the influence of extreme high-cost samples. Postprocessed random sampling is omitted for readability because its distribution is shifted relative to the main group. Lower values correspond to better portfolio solutions.
    }
    \label{fig:qpu_post_kde}
\end{figure}

Figures~\ref{fig:qpu_tail_kde} and~\ref{fig:qpu_post_kde} use feasible random samples as a stronger baseline.
These samples satisfy the portfolio constraints but are not optimized with respect to the objective.
Figure~\ref{fig:qpu_tail_kde} focuses on the final \(\alpha=0.1\) CVaR tail. Random sampling, bilinear, and HNDC-2 are omitted from this plot for readability because their distributions are shifted relative to the main group. In this tail comparison, HNDC-1 concentrates more probability mass at lower objective values than the feasible random baseline, while DGC shows a wider distribution with competitive low-cost samples.

Figure~\ref{fig:qpu_post_kde} shows the corresponding distributions after bit-flip postprocessing. The postprocessing step improves the distributions for all layouts, as expected from a local refinement
procedure. HNDC-1 remains the strongest performer after postprocessing and continues to shift probability mass below the feasible-random baseline. This suggests that, for this instance, HNDC not only helps the circuit reach feasible regions, but also improves the quality of feasible samples relative to random feasible
sampling.

\begin{figure}[t]
    \centering
    \includegraphics[width=0.5\columnwidth]{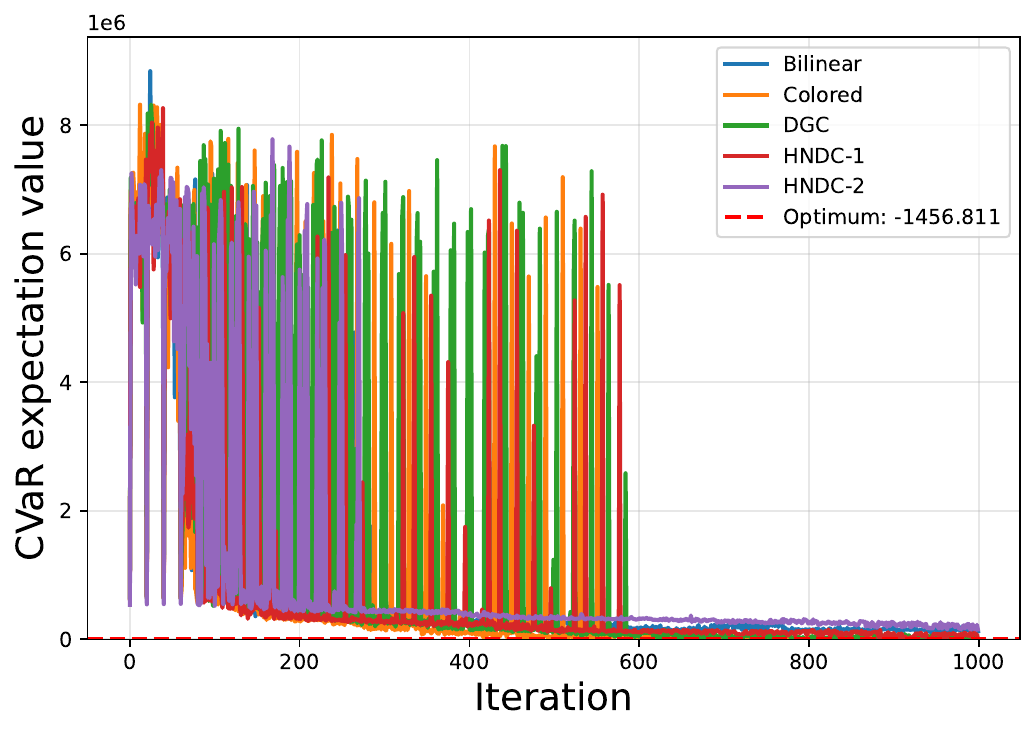}
    \caption{
    Evolution of the CVaR objective during QPU optimization for the tested ansatz layouts. The generally
    decreasing trajectories indicate improvement in the low-cost tail of the sampled distribution during
    parameter optimization.
    }
    \label{fig:qpu_evolution}
\end{figure}

\section{Discussion}
\label{sec:discussion}

The experiments highlight that, for this dynamic portfolio instance, VQA performance depends strongly on choices beyond the problem formulation. In the simulator studies, the best-performing
configuration was determined not by any single factor, but by the interaction among sampling, classical optimization, and ansatz trainability. The CVaR study shows that low-tail optimization
can improve the best observed solutions, but only with sufficient sampling. The poor performance of \(\alpha=0.1\) without shot scaling is therefore a useful negative result: CVaR is not a free
improvement, since focusing on a smaller tail increases estimator sensitivity unless the shot budget is adjusted. The adaptive schedule did not beat the best fixed CVaR on the best-solution metric, but gave the strongest final low-tail CVaR while using fewer shots on average, supporting its interpretation as a distribution-shaping rather than best-sample strategy.

The optimizer study shows a similar metric-dependent distinction. PSO and PSO\(\rightarrow\)NFT perform comparably, with PSO slightly better in final CVaR and the two-stage schedule slightly better in best objective value. This suggests that global exploration matters for this instance, while local NFT refinement can still improve the best individual solution once a promising region is found. The two-stage schedule is therefore not universally preferable to PSO, but the results show that the exploration--refinement structure is effective under the fixed budget and ansatz used here.
The stronger performance of both PSO-based methods relative to NFT and COBYLA further indicates that initialization and global search remain important in this high-dimensional sampled landscape.

The depth study emphasizes the trainability cost of repetitions. Although deeper ansatz circuits are more expressive in principle, the one-repetition ansatz performed best under a fixed evaluation budget, with two repetitions close behind and three repetitions worst. This is consistent with the larger number of parameters reducing optimization efficiency when evaluations are limited.
Sequential growth did not improve over the fixed two-repetition ansatz, suggesting that freezing earlier layers restricts useful co-adaptation. These observations are specific to the layout, optimizer,
and budget used here, but they motivate evaluating depth increases rather than assuming they help.

The QPU comparison provides the clearest evidence for the importance of hardware-aware structure. The bilinear layout is shallow before transpilation, but its two-qubit depth and gate count grow substantially after routing on \texttt{ibm\_quebec}. In contrast, the colored, DGC, and HNDC layouts are built from hardware-compatible edges and largely preserve their depth after transpilation. HNDC-1 performs best across the final QPU metrics, suggesting that increasing two-qubit interaction depth can be beneficial, particularly when the added interactions are hardware-compatible and do not introduce substantial routing overhead. This contrasts with additional ansatz repetitions, which increased the parameter count and hindered training under a fixed budget in the simulator
study. The colored-versus-DGC comparison further suggests that problem-variable-to-qubit assignment can help: since DGC preserves the colored edge sets, its gain is attributable to the
data-guided assignment rather than to increased depth. Because this margin is small and based on a single instance and backend, it should be read as suggestive.

The random and feasible-random baselines are important for interpretation. The feasible space is
minuscule relative to the full binary space, so beating unrestricted random alone is a weak benchmark. The feasible-random comparison, which conditions on the main constraints, is more informative. The fact that the optimized layouts reach this regime, and that HNDC-1 exceeds feasible random in both final CVaR and best-sample objective, suggests that the circuits are biased toward better feasible portfolios rather than sampling randomly. This is not evidence of quantum advantage: the baselines characterize sample quality, while SCIP provides the certified classical optimum for the instance.

Overall, sampling-based VQA performance for this instance is shaped by multiple interacting design choices: sampling strategy, optimizer scheduling, ansatz structure, and hardware compatibility. The findings should be interpreted within the scope of the reported experiments, which use one portfolio dataset and one QPU optimization run per layout. The results provide empirical evidence for this setting, while broader conclusions require evaluation across additional datasets and repeated hardware runs.

\section{Conclusion}
\label{sec:conclusion}

In this work, we studied several design choices that affect the performance of sampling-based VQA for dynamic portfolio optimization. We considered a 150-qubit portfolio instance and evaluated the impact of CVaR aggregation, classical optimizer scheduling, ansatz repetition depth, and hardware-aware ansatz layout design.

First, we evaluated a specific adaptive CVaR schedule, in which the sampled tail is gradually tightened during optimization, and compared it against fixed-CVaR baselines. The adaptive schedule was competitive with fixed low-CVaR configurations while requiring fewer shots on average, and it showed favorable performance when the final sample distribution was evaluated using a common low-tail CVaR metric. Second, we evaluated a two-stage classical optimizer that begins with PSO for global exploration and then switches to NFT for local coordinate-wise refinement. In our experiments, this schedule outperformed COBYLA and NFT and achieved slightly better final best-objective performance than pure PSO, while PSO remained competitive on the final CVaR metric.

Third, we compared different ansatz repetition strategies. Under a fixed evaluation budget, lower ansatz repetitions tended to perform better, suggesting that the reduced number of variational parameters and more frequent effective parameter updates can be beneficial in this setting. Finally, we introduced and tested two hardware-aware layout modifications: a data-guided colored layout and a Heavy-Hex Native Deep-Chain layout. The hardware results provide preliminary evidence that problem-aware variable assignment can provide modest gains over a standard colored layout, and that increasing hardware-aligned two-qubit interaction depth can improve performance in this setting.

Taken together, these results indicate that VQA-based portfolio optimization depends not only on the problem formulation, but also on the interaction between sampling strategy, optimizer behavior, ansatz trainability, and hardware compatibility. Future work should evaluate the proposed strategies across more datasets, market scenarios, and problem sizes.

\section*{Acknowledgments}
The author gratefully thanks Ibrahim Shehzad and Sean Wagner from IBM for their support throughout this project and for valuable technical discussions and insights. The author also acknowledges access to the \texttt{ibm\_quebec} quantum system for the hardware experiments through PINQ\textsuperscript{2}.

\appendix
\section{Data-Guided Colored Mapping}
\label{app:dgc_mapping}

This appendix gives the implementation details of the DGC. The goal is to choose a variable-to-qubit variable assignment that places strongly dependent logical variables on qubits connected by the fixed three-color entangling schedule.

\subsection{Hardware graph and mapping objective}

Let \(N_q\) be the number of logical binary variables and let
\[
\mathcal{P}=\{p_1,\dots,p_{N_q}\}
\]
be the selected qubits. The hardware-compatible colored layout provides three edge sets
\[
\mathcal{E}^{(c)}\subseteq \mathcal{P}\times\mathcal{P},
\qquad c\in\{0,1,2\},
\]
where each \(\mathcal{E}^{(c)}\) is a matching. Hence, all two-qubit gates in the same color class can be executed in parallel. We denote the union of the three edge sets by
\[
\mathcal{E}=\mathcal{E}^{(0)}\cup\mathcal{E}^{(1)}\cup\mathcal{E}^{(2)}.
\]

A mapping is a bijection
\[
f:\mathcal{P}\rightarrow \{1,\dots,N_q\},
\]
where \(f(p)\) is the logical variable assigned to qubit \(p\). Given an affinity matrix
\(A\in\mathbb{R}^{N_q\times N_q}\), the mapping objective is
\begin{equation}
\label{eq:dgc_mapping_objective}
F(f)
=
\sum_{c=0}^{2}
\alpha_c
\sum_{(p,q)\in\mathcal{E}^{(c)}}
A_{f(p),f(q)}.
\end{equation}
Unless otherwise stated, we use equal color weights
\(\alpha_0=\alpha_1=\alpha_2=1\). This objective favors assignments in which high-affinity logical variables are placed on qubits that are entangled by the colored layout.

\subsection{Sampling low-energy candidate solutions}

To estimate dependencies between logical variables, we first generate a pool of candidate bitstrings
\[
\mathcal{S}=\{z^{(s)}\}_{s=1}^{N},
\qquad z^{(s)}\in\{0,1\}^{N_q},
\]
using a classical heuristic sampler applied to the same portfolio objective \(E(z)\) used in the VQA evaluation. The sampler is used only to estimate a dependency structure; it is not used to solve the final VQA optimization problem.

To emphasize low-energy structure while keeping some diversity, we optionally assign Boltzmann-type weights to the samples:
\begin{equation}
\label{eq:dgc_sample_weights}
w_s
=
\frac{
\exp\left(
-\beta \frac{E(z^{(s)})-E_{\min}}{\sigma_E+\epsilon}
\right)
}{
\sum_{\ell=1}^{N}
\exp\left(
-\beta \frac{E(z^{(\ell)})-E_{\min}}{\sigma_E+\epsilon}
\right)
},
\end{equation}
where \(E_{\min}=\min_s E(z^{(s)})\), \(\sigma_E\) is the sample standard deviation of the energies, \(\beta\ge 0\) controls the emphasis on low-energy samples, and \(\epsilon>0\) is a small numerical stabilizer. If no reweighting is used, we set \(w_s=1/N\).

\subsection{Weighted mutual information}

From the weighted sample pool, we estimate a symmetric dependency matrix
\(W\in\mathbb{R}^{N_q\times N_q}\) using weighted mutual information. For variables \(i\) and \(j\),
\[
W_{ij}=I(z_i;z_j),
\]
where
\begin{equation}
\label{eq:dgc_mi}
I(z_i;z_j)
=
\sum_{a,b\in\{0,1\}}
p_{ij}(a,b)
\log
\frac{p_{ij}(a,b)}{p_i(a)p_j(b)}.
\end{equation}
The weighted empirical probabilities are estimated as
\[
p_i(a)
=
\sum_{s=1}^{N}
w_s\,\mathbf{1}\{z_i^{(s)}=a\},
\]
and
\[
p_{ij}(a,b)
=
\sum_{s=1}^{N}
w_s\,\mathbf{1}\{z_i^{(s)}=a,\;z_j^{(s)}=b\}.
\]
In practice, a small pseudocount is added to all empirical probabilities before normalization to avoid instability when some joint outcomes are rare. We set \(W_{ii}=0\) and symmetrize \(W\).

We also compute the weighted entropy of each variable:
\begin{equation}
\label{eq:dgc_entropy}
H_i
=
-\sum_{a\in\{0,1\}}
p_i(a)\log p_i(a).
\end{equation}
Variables with very small entropy are nearly frozen in the low-energy sample distribution, while variables with larger entropy remain more active.

\subsection{Softened affinity matrix}

Raw mutual information can be heavy-tailed, and aggressive entropy weighting may remove useful weak interactions when many variables are nearly frozen. We therefore define a softened affinity matrix
\begin{equation}
\label{eq:dgc_soft_affinity}
A_{ij}
=
W_{ij}
\left(
\eta+(1-\eta)\sqrt{H_iH_j}
\right),
\qquad \eta\in[0,1].
\end{equation}
The parameter \(\eta\) acts as a floor. When \(\eta=1\), the affinity reduces to the mutual-information
matrix \(A=W\). When \(\eta=0\), the affinity is fully entropy-weighted. Intermediate values preserve weak interactions while still reducing the influence of nearly frozen variables. We set \(A_{ii}=0\) and symmetrize \(A\).

\subsection{Swap-based local search}

The mapping problem in \eqref{eq:dgc_mapping_objective} is a sparse quadratic-assignment-like problem. We solve it using a swap-based local search. The search starts from an initialization that maps high-degree qubits to high-centrality logical variables, where logical centrality is computed as
\[
c_i=\sum_{j=1}^{N_q} A_{ij}.
\]
Then, at each iteration, the assignments of two qubits \(p,q\in\mathcal{P}\) are swapped:
\[
f'(p)=f(q),\qquad f'(q)=f(p),\qquad f'(r)=f(r)\;\;\forall r\notin\{p,q\}.
\]

The objective change can be evaluated locally. Let \(N(p)\) denote the neighbors of \(p\) in the union graph \((\mathcal{P},\mathcal{E})\). Since only edges incident to \(p\) or \(q\) can change, the difference
\(\Delta F=F(f')-F(f)\) is
\begin{align}
\Delta F
&=
\sum_{x\in N(p)\setminus\{q\}}
\left(
A_{f'(p),f(x)}-A_{f(p),f(x)}
\right)
\nonumber\\
&\quad+
\sum_{y\in N(q)\setminus\{p\}}
\left(
A_{f'(q),f(y)}-A_{f(q),f(y)}
\right)
\nonumber\\
&\quad+
\mathbf{1}\{(p,q)\in\mathcal{E}\}
\left(
A_{f'(p),f'(q)}-A_{f(p),f(q)}
\right).
\label{eq:dgc_delta}
\end{align}
Because the hardware graph has bounded degree, this local update is inexpensive.

We use a greedy acceptance rule: a proposed swap is accepted if \(\Delta F>0\). To improve robustness, we use multiple random restarts and retain the mapping with the largest value of \(F(f)\).

\subsection{Algorithm summary}
Algorithm~\ref{alg:dgc_mapping} summarizes this approach.

\begin{algorithm}[t]
\caption{Data-Guided Colored Mapping}
\label{alg:dgc_mapping}
\begin{algorithmic}[1]
\REQUIRE Candidate samples \(\{z^{(s)}\}_{s=1}^{N}\), energies \(E(z^{(s)})\), edge sets
\(\{\mathcal{E}^{(c)}\}_{c=0}^{2}\), softening parameter \(\eta\), restarts \(R\), iterations \(T\)
\STATE Compute sample weights \(w_s\) using Eq.~\eqref{eq:dgc_sample_weights}, or set \(w_s=1/N\)
\STATE Estimate weighted mutual information matrix \(W\) using Eq.~\eqref{eq:dgc_mi}
\STATE Estimate variable entropies \(H_i\) using Eq.~\eqref{eq:dgc_entropy}
\STATE Compute affinity matrix \(A\) using Eq.~\eqref{eq:dgc_soft_affinity}
\STATE Build the union graph \((\mathcal{P},\mathcal{E})\), where \(\mathcal{E}=\cup_c\mathcal{E}^{(c)}\)
\STATE \(f^\star \leftarrow \emptyset\), \(F^\star\leftarrow -\infty\)
\FOR{\(r=1,\dots,R\)}
    \STATE Initialize \(f\) using qubit degree and logical centrality
    \FOR{\(t=1,\dots,T\)}
        \STATE Propose candidate swaps of qubit assignments
        \STATE Compute \(\Delta F\) for each proposed swap using Eq.~\eqref{eq:dgc_delta}
        \STATE Apply the best swap if its \(\Delta F>0\)
    \ENDFOR
    \STATE Compute \(F(f)\) using Eq.~\eqref{eq:dgc_mapping_objective}
    \IF{\(F(f)>F^\star\)}
        \STATE \(f^\star\leftarrow f\), \(F^\star\leftarrow F(f)\)
    \ENDIF
\ENDFOR
\RETURN \(f^\star\)
\end{algorithmic}
\end{algorithm}

\section{Penalty Calibration via Objective Scaling}
\label{app:penalty_cal}

This appendix describes how we choose penalty magnitudes for the soft constraints in a way that is consistent across instances and avoids per-experiment tuning. We compute a conservative objective scale \(S_{\mathrm{obj}}\) from upper bounds on the main objective contributions and set the soft-constraint penalty coefficients proportional to this scale. The resulting penalties are fixed for a given problem instance and are used unchanged across all VQA configurations.

\subsection{Objective scale \(S_{\mathrm{obj}}\)}
Let \(p_t\in\mathbb{R}^n_{\ge 0}\) be the anchor prices at time \(t\), \(\Sigma_t\in\mathbb{R}^{n\times n}\) the within-period covariance matrix, and let \(H=h_{\max}\) be the maximum representable holding.
When a target cardinality \(B\) is provided, we tighten bounds using Top-\(B\) (or Top-\(2B\)) subsets.

For each period \(t=0,\dots,m-1\), we compute conservative upper bounds on the absolute contribution of each objective component used in \eqref{eq:dpo_obj}:

\paragraph{Return bound}
For \(t=0,\dots,m-1\), define \(\Delta p_t=p_{t+1}-p_t\) and
\((\Delta p_t)^+=\max(\Delta p_t,0)\) elementwise. We bound the magnitude of the return contribution by
\[
R_t^{\max} =
\begin{cases}
H \sum_{i\in \mathrm{Top}\text{-}B} (\Delta p_{i,t})^+, & 0<B<n,\\[2pt]
H \sum_{i=1}^n (\Delta p_{i,t})^+, & \text{otherwise.}
\end{cases}
\]

\paragraph{Risk bound}
If \(0<B<n\), let \(S\) be the indices of the \(B\) largest entries of \(p_t^2\), and denote by \(p_{t,S}\)
and \(\Sigma_{t,S}\) the corresponding restricted vector and submatrix. Then
\[
V_t^{\max} = \lambda H^2\, p_{t,S}^\top \Sigma_{t,S}\, p_{t,S}.
\]
Otherwise, using the spectral bound with \(\lambda_{\max}=\max|\mathrm{eig}(\Sigma_t)|\),
\[
V_t^{\max} = \lambda H^2\, \lambda_{\max}\, \|p_t\|_2^2.
\]

\paragraph{Transaction-cost bound.}
Using absolute changes in integer holdings, a conservative bound is
\[
T_t^{\max} =
\begin{cases}
\delta H \sum_{i\in \mathrm{Top}\text{-}k} p_{i,t}, & k=\min(2B,n),\;\; 0<B<n,\\[2pt]
\delta H \sum_{i=1}^n p_{i,t}, & \text{otherwise.}
\end{cases}
\]

\paragraph{Cash-term bound}
The cash interest term in \eqref{eq:dpo_obj} depends on the invested value \(V_t=\sum_i p_{i,t} h_{i,t}\).
We bound \(V_t\) by the maximum invested value achievable under the holding cap:
\[
I_t^{\max} = |\nu| \cdot
\begin{cases}
H \sum_{i\in \mathrm{Top}\text{-}B} p_{i,t}, & 0<B<n,\\[2pt]
H \sum_{i=1}^n p_{i,t}, & \text{otherwise.}
\end{cases}
\]

Summing these per-period bounds yields
\[
S_{\mathrm{obj}} = \sum_{t=0}^{m-1} \big(|V_t^{\max}| + |R_t^{\max}| + |T_t^{\max}| + |I_t^{\max}|\big)
+ |L^{\max}|,
\]
where the liquidation-cost bound is
\[
L^{\max} =
\begin{cases}
\delta H \sum_{i\in \mathrm{Top}\text{-}B} p_{i,m}, & 0<B<n,\\[2pt]
\delta H \sum_{i=1}^n p_{i,m}, & \text{otherwise.}
\end{cases}
\]

\subsection{Penalty coefficients}
We scale penalties for soft constraints using \(S_{\mathrm{obj}}\):
\[
P_{\mathrm{card}} = \alpha_{\mathrm{card}}\, S_{\mathrm{obj}},\qquad
P_{\mathrm{hold}} = \alpha_{\mathrm{hold}}\, S_{\mathrm{obj}},
\]
where \(\alpha_{\mathrm{card}},\alpha_{\mathrm{hold}}>0\) are fixed dimensionless multipliers (shared across experiments). The cardinality penalty targets a desired number of active assets per period (penalizing \(|a_t-B|\)), while the holding penalty discourages any \(h_{i,t}>h_{\max}\). 

\bibliographystyle{unsrt}
\bibliography{references}

\end{document}